\documentclass[twocolumn, floatfix]{aastex701}
\usepackage{amsmath} 

\usepackage{comment}
\usepackage{appendix}
\usepackage{multirow}

\providecommand{\sorthelp}[1]{}
\defcitealias{cordova_rosado_cross-correlation_2024a}{CR24a}
\defcitealias{cordova_rosado_cross-correlation_2024b}{CR24b}

\begin{document}

\title{Cross-correlation of Luminous Red Galaxies with ML-selected AGN in HSC-SSP III: \\ HOD Parameters for Type I and Type II Quasars}

\shorttitle{HOD Parameters for Type I and Type II Quasars}
\shortauthors{C\'ordova~Rosado et al.}

\author[0000-0002-7967-7676]{Rodrigo~C\'ordova~Rosado}
\email{rodrigo.cordova_rosado@cfa.harvard.edu}
\affiliation{Department of Astrophysical Sciences, Peyton Hall, Princeton University, 4 Ivy Lane, Princeton, NJ 08544, USA}
\affiliation{Center for Astrophysics $|$ Harvard \& Smithsonian, 60 Garden Street, Cambridge, MA, 02138, USA}

\author[0000-0003-4700-663X]{Andy~D.~Goulding}
\email{goulding@astro.princeton.edu}
\affiliation{Department of Astrophysical Sciences, Peyton Hall, Princeton University, 4 Ivy Lane, Princeton, NJ 08544, USA}

\author[0000-0002-5612-3427]{Jenny~E.~Greene}
\email{jgreene@astro.princeton.edu}
\affiliation{Department of Astrophysical Sciences, Peyton Hall, Princeton University, 4 Ivy Lane, Princeton, NJ 08544, USA}

\author[0000-0002-5808-4708]{Nickolas~Kokron}
\email{kokron@ias.edu}
\affiliation{Department of Astrophysical Sciences, Peyton Hall, Princeton University, 4 Ivy Lane, Princeton, NJ 08544, USA}
\affiliation{School of Natural Sciences, Institute for Advanced Study, 1 Einstein Drive, Princeton, NJ, 08540, USA}

\author[0000-0003-2792-6252]{Andrina~Nicola}
\email{andrina.nicola@manchester.ac.uk}
\affiliation{Jodrell Bank Centre for Astrophysics, Department of Physics and Astronomy, The University of Manchester, Manchester M13 9PL, UK}

\author[0000-0002-0106-7755]{Michael~A.~Strauss}
\email{strauss@astro.princeton.edu}
\affiliation{Department of Astrophysical Sciences, Peyton Hall, Princeton University, 4 Ivy Lane, Princeton, NJ 08544, USA}

\author[0000-0003-1468-9526]{Ryan~C.~Hickox}
\email{Ryan.C.Hickox@dartmouth.edu}
\affiliation{Department of Physics and Astronomy, Dartmouth College, 6127 Wilder Laboratory, Hanover, NH 03755, USA}

\correspondingauthor{Rodrigo~C\'ordova~Rosado}
\email{rodrigo.cordova\_rosado@cfa.harvard.edu}

\begin{abstract}

Understanding the dark matter (DM) halo environment in which galaxies that host active galactic nuclei (AGN) reside is a window into the nature of supermassive black hole (SMBH) accretion. We apply halo occupation distribution (HOD) modeling tools to interpret the angular cross-correlation functions between $1.5\times10^6$ luminous red galaxies (LRGs) and our $\sim28,500$ Hyper Suprime-Cam + Wide-field Infrared Survey Explorer-selected (and $L_{6 \mu m}$-limited) AGN to infer the halo properties of distinct quasar samples at physical scales $s>0.1\,{\rm Mpc}$, for $z\in0.7-1.0$. We find that Type I (unobscured) and Type II (obscured) AGN cluster differently, both on small and large physical scales. The derived HODs imply that Type I AGN reside, on average, in substantially ($\sim3\times$) more massive halos ($M_h \sim 10^{13.4} M_\odot$) than Type II AGN ($M_h \sim 10^{12.9} M_\odot$) at $>5\sigma$ significance. While Type II AGN show one-halo correlations similar to that of galaxies of their average halo mass, the Type I AGN intra-halo clustering signal is significantly shallower. We interpret this observation with HOD methods and find Type I AGN are significantly less likely ($f_{sat}\sim0.05^{+1}_{-0.05}\%$) to be found in satellite galaxies than Type II AGN. We find reddened + obscured AGN to have typical satellite fractions for their inferred average halo mass ($\sim10^{13} M_\odot$), with $f_{sat} \sim 20^{+10}_{-5}\%$. Taken together, these results pose a significant challenge to the strict unified AGN morphological model, and instead suggest that a quasar's spectral class is strongly correlated with its host galaxy's dark matter halo environment. These intriguing results have provided a more complex picture of the SMBH -- DM halo connection, and motivate future analyses of the intrinsic galaxy and accretion properties of AGN.

\end{abstract}


\section{Introduction}

The growth of supermassive back holes (SMBHs) is an essential aspect of galaxy evolution, with significant influence on the overall properties of the galaxies in which they reside \citep{kormendy_inward_1995, kormendy_coevolution_2013}. When SMBHs undergo rapid matter accretion, their accretion disks outshine all other luminous matter, providing a clear target to study SMBH growth as an active galactic nucleus \citep[AGN, ][]{schmidt_3c_1963}. Studying the interplay between AGN activity and its connection to large scale structure is crucial to understand SMBH-galaxy co-evolution \citep{fabian_observational_2012, kormendy_coevolution_2013, heckman_coevolution_2014}. As we continue exploring the high redshift universe, it is essential we build a robust understanding of the evolution and mechanisms of AGN occurrence \citep{fan_quasars_2023, harikane_jwstnirspec_2023, greene_uncover_2024}.

AGN have been historically split into unobscured (Type I) and obscured (Type II) classes, based on the level of reddening, or the presence of broadened emission lines, in the optical-UV spectrum \citep[e.g.,][]{netzer_revisiting_2015, hickox_obscured_2018, alexander_what_2025}. The unified AGN model proposes that these differences in obscuration are mediated by a dusty flattened region on parsec scales from the accretion disk, described as a torus, that acts as a screen of the broad line region when observed from particular inclinations \citep{antonucci_unified_1993, urry_unified_1995, netzer_revisiting_2015}. Under unification, AGN of different classes are intrinsically the same class of object, being triggered at potentially different stages of galactic growth, but with no correlation between the degree of obscuration and other inferred properties like black hole, stellar, or halo mass \citep{almeida_nuclear_2017}. And yet, there have been significant challenges to the strict unified explanation of AGN phenomenology. Instead, these spectral types have been thought to be possible markers of different evolutionary steps of AGN formation histories \citep{hopkins_cosmological_2008, hickox_host_2009}. Prior studies have shown that there are potential correlations between obscuration effects and the overall markers of galaxy evolutionary stage \citep{sanders_ultraluminous_1988, canalizo_quasi-stellar_2001, hickox_clustering_2011, allevato_clustering_2014, fawcett_striking_2023, petter_host_2023}. Other studies have shown evidence that obscured AGN are more likely to be part of a galaxy merger system than unobscured AGN \citep{mihos_gasdynamics_1996, blain_dust-obscured_1999, urrutia_evidence_2008, koss_host_2011, ellison_galaxy_2011, ellison_galaxy_2013, ellison_definitive_2019, glikman_major_2015, goulding_galaxy_2018, secrest_x-ray_2020, ricci_hard_2021}. The relative impact on total obscuration originating from the torus versus galaxy-scale dust also remains unclear \citep{goulding_towards_2009, goulding_deep_2012}. Taken together, these results substantively challenge a unified AGN model as a capacious framework with which to understand the source of different AGN spectral types. 

However, analyzing these trends for single objects is highly degenerate with the particular properties of any one system and its host galaxy. Thus, galaxy clustering statistics have become a standard approach to distinguish population-level characteristics from individual AGN observations, and to infer properties about the dark matter (DM) environment that host AGN \citep{osmer_three-dimensional_1981, shaver_clustering_1984, shanks_spatial_1987, iovino_clustering_1988, andreani_evolution_1992, mo_quasar_1993, shanks_qso_1994, croom_qso_1996, la_franca_quasar_1998,  croom_2df_2005, lidz_luminosity_2006, shen_biases_2008, cappelluti_active_2010, cappelluti_clustering_2012, shen_cross-correlation_2013, eftekharzadeh_clustering_2015,mendez_primus_2016, laurent_clustering_2017,  toba_clustering_2017, he_clustering_2018,chaussidon_angular_2022, arita_subaru_2023, krumpe_spatial_2023, eilers_eiger_2024, pizzati_unified_2024}. The wide range of results have not converged on whether Type I or Type II AGN are hosted by more massive DM halos, or if there is any statistically significant difference \citep{hickox_clustering_2011, allevato_clustering_2014, dipompeo_angular_2014, dipompeo_updated_2016, jiang_differences_2016, dipompeo_characteristic_2017, koutoulidis_dependence_2018, powell_swiftbat_2018, petter_host_2023, li_black_2024}. However, a substantial fraction of these studies have been hitherto constrained by their sample size, area, redshift availability, and other factors that limit one's ability to measure a robust correlation function for AGN sub-types \citep[cf.][]{coil_aegis_2009, gilli_spatial_2009, cappelluti_active_2010, allevato_xmm-newton_2011, koutoulidis_clustering_2013,  krumpe_spatial_2018, petter_host_2023}.

As we have shown in \citet{cordova_rosado_cross-correlation_2024a, cordova_rosado_cross-correlation_2024b} (hereafter \citetalias{cordova_rosado_cross-correlation_2024a}, \citetalias{cordova_rosado_cross-correlation_2024b}), we attempted to resolve several of these systematic effects that could potentially lead to inconsistent results across datasets. In \citetalias{cordova_rosado_cross-correlation_2024a}, we presented the spatial correlations of Type I and Type II AGN identified with an unsupervised machine-learning selection combining Hyper Suprime-Camera \citep[HSC, ][]{aihara_hyper_2018} optical and Wide-field Infrared Explorer \citep[WISE, ][]{wright_wide-field_2010} mid-infrared (MIR) photometry. In \citetalias{cordova_rosado_cross-correlation_2024b}, we matched the HSC+$WISE$-selected AGN with Dark Energy Spectroscopic Instrument \citep[DESI, ][]{desi_collaboration_desi_2016, desi_collaboration_early_2024} early release data to mitigate concerns over photometric redshift uncertainties in our correlation analyses. In both, we showed for a redshift and luminosity-constrained sample, that the linear clustering amplitude (often referred to as the two-halo term - i.e. clustering at large scales between halos) of unobscured AGN was substantially ($\sim 5\times$) larger than that of obscured AGN.

Having previously measured the two-halo term, we turn our attention to the smaller scale -- and intra-halo clustering driven -- one-halo term. Given its rising correlation at small scales, it affords us the highest signal-to-noise (S/N) measurements of the AGN clustering statistic. Using standard Halo Occupation Distribution \citep[HOD, ][]{cooray_halo_2002, berlind_halo_2002, zheng_theoretical_2005, zheng_galaxy_2007, zehavi_galaxy_2011, nicola_tomographic_2020} tools, we are able to tie the amplitude and shape of the one-halo term to physically-interpretable quantities, including the satellite fraction -- the proportion of galaxies in our sample that are satellites in a halo. 

Many prior studies have performed HOD analyses of Type I AGN \citep{miyaji_spatial_2010, starikova_constraining_2011, kayo_very_2012, allevato_occupation_2012, richardson_halo_2012, shen_cross-correlation_2013, jiang_differences_2016, krumpe_spatial_2018, powell_swiftbat_2018, eftekharzadeh_halo_2019, krumpe_spatial_2023}, consistently converging on a low satellite fraction relative to other galaxies of their typical average halo mass. These studies have also often inferred that Type II AGN could be found in satellites more often than Type I's \citep{villarroel_different_2014, jiang_differences_2016, krumpe_spatial_2018, powell_swiftbat_2018}. However these studies are often conducted at low redshift ($z<0.2$), with poorer spatial resolution, and/or smaller areas. These, along with selection effects and uneven coverage in redshift and luminosity, have led to significant questions on the reliability of described halo differences for AGN \citep[see also ][]{aird_agn-galaxy-halo_2021}. Moreover, varying the choice of HOD parameterization gives inconsistent values of $f_{sat}$, driven by the number of HOD parameters -- which can change the models' flexibility in the transition from the two-halo to one-halo components. With these considerations in mind, we use the HOD model to interpret the clustering of our distinct AGN samples.

This paper is organized as follows. In \S \ref{sec:data}, we summarize the datasets used in \citetalias{cordova_rosado_cross-correlation_2024a} that are reused here. In \S \ref{sec:methods}, we outline our methods to measure the projected angular correlation function and its uncertainties, to define and fit the HOD to the clustering signal, and to infer its derived parameters. We show the results of our correlation functions and their HOD model fits in \S \ref{sec:results}. We discuss our results and their implications for AGN halo environments in \S \ref{sec:disc}, and summarize our conclusions in \S \ref{sec:conclu}.

In this work, we will use the measured angular correlation functions from \citetalias{cordova_rosado_cross-correlation_2024a}, and focus on measuring the average intra-halo properties as encoded in the one-halo term. Comparing our fitted parameters with alternative HOD parametrizations, we investigate if there are any inherent differences in small-scale clustering of different AGN spectral classes, given the differences we have found on large clustering scales. Throughout this analysis, we adopt a ``Planck 2018''  $\Lambda$CDM cosmology \citep{planck_collaboration_planck_2020}, with $h = H_0/100\,{\rm km\, s^{-1} Mpc^{-1}} = 0.67$, $\Omega_c = 0.1198/h^2 = 0.267$, $\Omega_b = 0.02233/h^2 = 0.0497$, $n_s = 0.9652$, and $\sigma_8 = 0.8101$, quoting parameters in ``$h$-less'' units, so as to be in line with recent cosmological clustering analyses \citep[e.g.][]{nicola_tomographic_2020}. Quantities defined with a $\log$ are exclusively $\log_{10}$ values, and we use $\ln$ to indicate our use of the natural logarithm. Normalized distributions are defined such that the integral of the distribution is equal to unity. We express photometric magnitudes in the AB system \citep{oke_secondary_1983}. In the context of galaxy bias and halo mass parametrization, we use the \cite{tinker_large-scale_2010} formalism with $\Delta_m = 200$ (the spherical overdensity radius definition) for the halo mass function, i.e. $M_{200}$. Foreground dust extinction has been corrected in all observations as supplied in the HSC catalog \citep{aihara_third_2022} based on \cite{schlegel_maps_1998}.

\section{Data}\label{sec:data}

\subsection{HSC Photometry and Galaxy Catalog}

The Hyper Suprime-Cam Strategic Strategic Program \citep[HSC-SSP, ][]{aihara_hyper_2018} uses its namesake wide-field camera on the 8.2m Subaru Telescope atop Maunakea, Hawai'i to study galactic history with \textit{grizy} wide-band photometry, across 330 nights of observations. Leveraging the 1.77\,deg$^2$ field of view and deep photometric sensitivity, HSC-SSP produced a 670\,deg$^2$ full-depth and full-color Wide Survey as part of its most recent public data release \citep[PDR3, ][]{aihara_third_2022}. We make use of the galaxy catalog as described in \citetalias{cordova_rosado_cross-correlation_2024a}, and review salient details here. As discussed in \cite{aihara_third_2022}, the HSC-SSP PDR3 data are released with a full-depth-full-color mask, as well as a bright star mask, detailed in \cite{coupon_bright-star_2018}. After removing areas with bright sources identified with the Gaia DR2 bright star catalog \citep{gaia_collaboration_gaia_2018}, we additionally apply a source mask based on \textit{WISE} imaging data \citep{wright_wide-field_2010, cutri_vizier_2012}. We make use of a magnitude-limited ($i<24$) galaxy sample to identify both our luminous red galaxy (LRG) and AGN sources. Our experiments are carried out on the three largest HSC PDR3 equatorial fields with full coverage: XMM-LSS (hereafter XMM), VVDS, and GAMA. Herein, we select the most robust photo-$z$ catalog possible, identified by implementing a color-color cut in $g-r > 1.2$ and $r-i<1.0$ to select the LRGs in the catalog. The details and motivation of this selection are explored in Appendix A of \citetalias{cordova_rosado_cross-correlation_2024a}. The area, number of objects (including the number of spectroscopic vs. photometric redshifts), and location of the HSC-SSP fields are detailed in \S 2 of \citetalias{cordova_rosado_cross-correlation_2024a}, and key details are shown in Table \ref{tab:lrg_agn_n}. These LRG samples will be cross-correlated with the AGN identified from HSC and \textit{WISE} photometry, which are described in the following subsection.

\begin{table}
\centering
\caption{Field properties and number of objects with $>3\%$ of their $p(z)$ in our redshift range ($z\in 0.7-1.0$). }
\begin{tabular}{c|ccc}

\hline \hline  

&GAMA & VVDS & XMM\\
\hline
Area [deg$^2$]  & 397.18  &  100.95 &70.42 \\

$N_{obj}$ (LRGs)  & 1,054,791 &  271,873  & 183,241  \\

$N_{obj}$ (AGN)$^{*\dagger}$  & 19,442 & 4,540  &  4,512 \\

$N_{obj}$ (unobscured AGN)$^*$  & 6,510 & 863  & 893  \\

$N_{obj}$ (reddened AGN)$^*$  & 5,100 & 920  & 870  \\

$N_{obj}$ (obscured AGN)$^*$  & 7,391  & 2,697   & 2,697  \\ 

\hline \hline 

\end{tabular} 

\raggedright
\vspace{0.05in}
\footnotesize
      $^*$ Luminous AGN sample ($L_{6\mu m} > 3\times 10^{44}$ erg s$^{-1}$)
      
      $^\dagger$ There is a small fraction of AGN that do not have a confident spectral classification, such that the subsamples presented here do not sum to the total number of AGN selected using unsupervised machine-learning classification tools.
    \label{tab:lrg_agn_n} 
\end{table}

\subsection{Optical and MIR-selected AGN in HSC and \textit{WISE}}\label{sec:dat_AGN}

A key improvement over previous quasar clustering analyses is our substantially increased number density of AGN, a result of the combination of deep optical imaging from HSC, \textit{WISE} \citep{wright_wide-field_2010} mid-IR photometry, and unsupervised machine learning classification techniques. We describe the identification and classification process for the AGN samples used in this analysis in \S 2.4 of \citetalias{cordova_rosado_cross-correlation_2024a}. The joint HSC-SSP and \textit{WISE} AGN catalog, hereafter HSC+\textit{WISE}, is detailed in Goulding et al. (in-prep.) and contains $>340$ deg$^{-2}$ AGN across the HSC Wide survey fields. To create this catalog, we match \textit{grizy} photometry from HSC-SSP to WISE sources with S/N$>5$ in the \textit{W1} band photometry in all\textit{WISE} and un\textit{WISE} MIR catalogs \citep{cutri_vizier_2012, schlafly_unwise_2019} using a maximum likelihood estimator. Combining source catalogs, we require that each source have a detection threshold of S/N$>$ 4,3,3 in their \textit{g}, $W2$, and $W3$ observations, respectively. 

As detailed in \S 2.4.1 of \citetalias{cordova_rosado_cross-correlation_2024a}, the objects are probabilistically classified as either unobscured, reddened, and obscured AGN based on their redshift and $g-W3$ color. Unobscured AGN have strong UV components and are presumed to have broad emission lines based on their photometry, i.e. to be Type I AGN, given the spectroscopic training data. Reddened objects, meanwhile, are selected based on the similarity of their SEDs to AGN for whom we spectroscopically confirm broad line emission, but with significant optical-UV reddening from dust. These have also been described as ``red quasars'' in the literature \citep{fawcett_striking_2023}. Lastly, the obscured AGN sample is identified for its similarity to Type II AGN, characterized by their narrow line emission (and lack of broad lines) and significant dust obscuration in the optical-UV. Additionally, we impose the same luminosity limit we have previously used, requiring that the considered AGN have $L_{6 \mu m} > 3 \times 10^{44}$ erg s$^{-1}$. The luminosity distributions for the different subsamples are consistent upon implementing this cut, as shown in Figure 4 of \citetalias{cordova_rosado_cross-correlation_2024a}. 

\cite{hviding_spectroscopic_2024} showed (for objects brighter than $i<22.5$) that our photometrically-classified obscured and unobscured objects were consistently confirmed as AGN via spectroscopic follow-up. In \citetalias{cordova_rosado_cross-correlation_2024b}, we further verified AGN identification from HSC and WISE photometry by matching HSC-identified sources with the Dark Energy Spectroscopic Instrument's \citep[DESI, ][]{desi_collaboration_desi_2016} early data release spectra \citep{desi_collaboration_early_2024}. In doing so, we showed that the total inferred number density of confirmed AGN in our sample is at least $60\%$ higher than that of previous optical-MIR selections \citep[cf.][]{assef_wise_2018}, as detailed in \S 4 of \citetalias{cordova_rosado_cross-correlation_2024b}. This increase is driven by a significant improvement in the identification and confirmation of obscured AGN sources.

\section{Methodology}\label{sec:methods}
\begin{figure*}
    \centering
    \includegraphics[width=1.\linewidth]{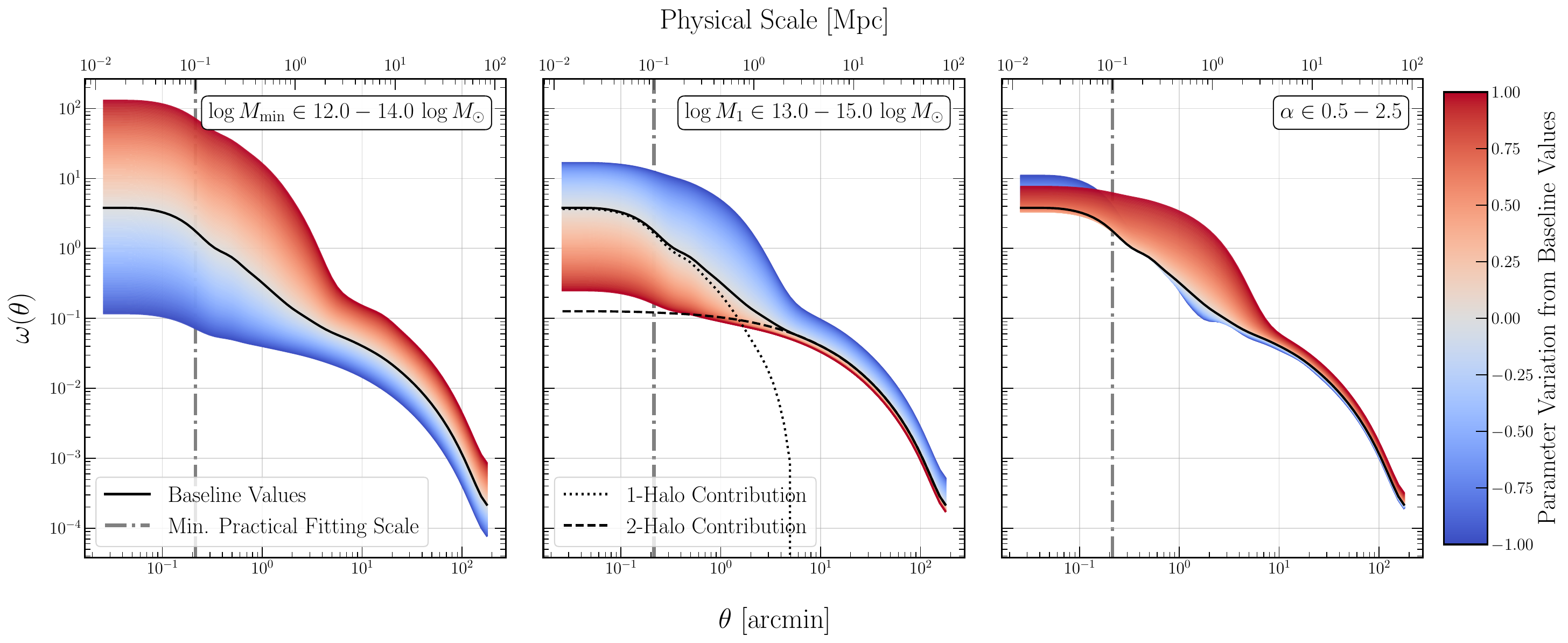}
    \caption{Pedagogical example of the effect of changing the values of the 3-parameter HOD model that we project to its angular correlation function ($\omega(\theta)$) form, as outlined in \S \ref{sec:HOD_method}. The baseline values are $\log M_{\rm min} = 13 \log M_\odot$, $\log M_{1} = 14 \log M_\odot$, $\alpha = 1.5$. The physical scales are converted assuming $z=0.8$. The axes are shared between panels, the middle panel shows the relative contributions of the one- and two-halo terms, while the dotted-dashed line is the minimum scale we will fit data to with this model. We iteratively change one parameter by $\pm1.0$ in each panel, as indicated by the text in the upper right of each panel. The colored shading relates the shift in the parameter value to the model $\omega(\theta)$ it produces. $\log M_{\rm min}$ and $\log M_{1}$ have relatively straightforward (and opposite) effects on the amplitudes of the models, while $\alpha$ affects the one-halo term as a $\theta$-dependent scaling.}
    \label{fig:HOD_pedadogy}
\end{figure*}

We now describe the methods to measure the clustering statistic from our LRG and AGN catalogs, and interpret the contributions of the one-halo and two-halo terms using a Halo Occupation Distribution (HOD) model. 

\subsection{Clustering Measurements}\label{sec:clustering_methods}

\subsubsection{Angular Correlation Function}
The angular correlation function is the measurement of the excess probability of a pair of objects being separated by an angle $\theta$ above a Poisson distribution \citep{peebles_statistical_1973}. Spatial correlation statistics are essential tools with which to probe the clustering properties of galaxy populations and connect the empirical clustering to physically-motivated models of the galaxies dark matter environments, i.e. their halos. We seek to measure the intra-halo clustering from our AGN catalogs, necessitating a high density sample of quasars, which are presently only available in photometric catalogs like ours. We rely predominantly on photometric redshifts for our sample at $z\sim1$, leading to our analysis with the projected angular statistic, rather than a three-dimensional one. In order to reduce the impact of shot noise bias, we adopt the \cite{landy_bias_1993} estimator for the angular two-point function, defined for two distinct datasets $D_1$ \& $D_2$ as:
\begin{equation}
    \omega(\theta)=\frac{D_1 D_2(\theta) - D_1 R_2 (\theta) - D_2 R_1 (\theta) + R_1 R_2 (\theta) }{R_1 R_2 (\theta)},
\end{equation}
where this form reduces to the standard autocorrelation when $D_1 = D_2$, such that it does not require $R_1 \neq R_2$. 

As detailed in Appendix B of \citetalias{cordova_rosado_cross-correlation_2024a}, we employ a weighted clustering statistic to utilize the measurement of the full photometric redshift solution posterior, $p(z)$, of each object and its overlap with our redshift bin. We found that this weighted estimator is unbiased relative to other, tomographic, methods of measuring and modeling the linear components of the angular correlation function, but this method includes the probability of an object scattering into the redshift bin of interest where other methods do not. In \citetalias{cordova_rosado_cross-correlation_2024a}, we established 24 spatial bins for the angular correlation function estimation, logarithmically spaced from $s = 0.01 \, h^{-1} {\rm Mpc}$ to $s =100 \, h^{-1} {\rm Mpc}$. These physical bins are then converted into angular bins via the standard angular diameter distance conversion, with the median of each sample's $dN/dz$ used as the fiducial redshift to find the comoving distance. 

As defined in \citetalias{cordova_rosado_cross-correlation_2024a, cordova_rosado_cross-correlation_2024b}, we use a jackknife procedure to estimate the full covariance matrix for our clustering statistic measurements. This method estimates the statistical and field-level systematic uncertainty by iteratively removing 1/25 of the total area considered and recalculating the correlation statistic. The $1\sigma$ per-$\theta$-bin uncertainties represented as error bars throughout this work are drawn from the square root of the diagonal of the covariance matrix. Additionally, all reported values will be constrained from joint likelihood fits to the auto- and cross-correlations across the three fields, in turn folding in any field-to-field variability into the uncertainty estimation for the final set of fitted parameters (detailed in \S \ref{sec:fitting}).

\subsubsection{Redshift Bin}

We analyze the angular correlation function measurements we first presented in \citetalias{cordova_rosado_cross-correlation_2024a}, and refer the reader to \S 3.3 for details on the chosen redshift bins of the analysis, $z\in 0.7-1.0$.

\subsection{Clustering Interpretation}\label{sec:HOD_method}
Based in peak-background split theory and standard halo modeling approaches \citep{sheth_large_1999, cooray_halo_2002}, we can interpret the excess clustering relative to a dark matter distribution to measure bulk properties of the halos in which the galaxies in our samples reside with halo occupation distribution modeling tools. This assumes that all DM is essentially found within halos, and the HOD defines how galaxies populate these halos. These HOD models are fully implemented using the Core Cosmology Library \citep[CCL, ][]{chisari_core_2019}. 

\subsubsection{Halo Occupation Distribution Modeling}
Using a standard approach as outlined by \cite{zheng_galaxy_2007, zehavi_galaxy_2011}, we primarily employ the 3-parameter HOD consistently used in angular correlation function analyses as defined by \cite{seo_passive_2008}, where the total galaxy distribution is: 

\begin{equation}\label{eq:NM}
    \langle N(M) \rangle  = \langle N_c(M) \rangle \big(1 + \langle N_s (M)\rangle\big),
\end{equation}
and the central and satellite galaxy distribution terms are: 
\begin{equation}
     \langle N_c (M)\rangle = \exp \left(- \frac{M_{\rm min}}{M}\right),
\end{equation}
\begin{equation}\label{eq:Ns3}
    \langle N_s (M)\rangle = \left( \frac{M}{M_1}\right)^\alpha.
\end{equation}
These terms specify the mean number of galaxies (centrals or satellites) in a DM halo of mass $M$. The fitting parameters are:  $M_{\rm min}$, the truncation mass set for the central galaxy distribution (i.e. the minimum halo mass in which a central galaxy would be found); $M_1$, the mass at which one finds one satellite per halo; and $\alpha$, the slope of the power law describing the satellite number. The impact of changing different HOD parameters for our $\omega(\theta)$ model are visualized in Figure \ref{fig:HOD_pedadogy}. We choose this parameterization to be conservative with our number of free parameters for fitting angular correlation functions, following prior analyses like \cite{sawangwit_angular_2011}. 

These halo occupation definitions are then passed into the definition of the galaxy-galaxy power spectrum, which is split into its one-halo and two-halo components,
\begin{equation} \label{eq:Pgg_def}
    P_{gg} (k,z) = P_{gg}^{1h}(k, z) + P_{gg}^{2h}(k,z).
\end{equation}
For an autocorrelation, the one-halo term is defined as
\begin{multline}
    P_{gg}^{1h}(k,z) = \frac{1}{\bar{n}_g(z)^2}\int dM\,  n(M,z) \, u(k|M)\, \langle N_c(M) \rangle \\ \times \Big[ \langle N_c(M) \rangle \langle N_s^2 (M)\rangle\, u(k|M) + 2\,\langle N_s (M)\rangle \Big],
\end{multline}
where 
\begin{equation}
    \bar{n}_g (z) = \int dM \, n(M, z)\, \langle N(M) \rangle,
\end{equation}
$n(M,z)$ is the halo mass function as defined in \cite{tinker_large-scale_2010} for $\Delta_m = 200$, and $u(k|M)$ is the Fourier transform of the normalized density profile for halos \citep[i.e. an NFW profile, ][]{navarro_universal_1997}. Satellites are assumed to trace the DM distribution and are thus modeled with an NFW. We use the DM halo concentration from \cite{duffy_dark_2008} when defining the NFW profile. Meanwhile, the two-halo term for a galaxy autocorrelation is 
\begin{multline}
    P_{gg}^{2h}(k,z) =\Bigg( \frac{1}{\bar{n}_{g}(z)}\int dM\,  n(M,z) \, \langle N_c(M) \rangle \times \\ \Big[ 1 + \,\langle N_s (M)\rangle\, u(k|M) \Big] \Bigg)^2 P_{\rm lin}(k,z), 
\end{multline}
where $P_{\rm lin}(k)$ is the linear matter power spectrum. 

We follow \cite{krolewski_tomographic_2025} to implement the cross-correlation calculation and include all necessary (cross) terms for the one-halo term:
\begin{multline}
    P^{1h}_{g_1 g_2} (k,z) =  \frac{1}{\bar{n}_{g_1} \,\bar{n}_{g_2}  } \int dM\, n(M,z) \, u(k|M) \, \langle N_{c,1} N_{c,2} \rangle \\ \times  \Big[\langle N_{s,1} N_{s,2} \rangle \, u(k|M) \, + \,  \langle N_{s,2} \rangle \, + \,  \langle N_{s,1} \rangle \Big],
\end{multline}
where the numbered subscripts denote the two samples being cross-correlated. Similarly for the two-halo term, we define 
{
\begin{multline}
    P_{g_1 g_2}^{2h}(k,z) = \frac{1}{\bar{n}_{g_1} \,\bar{n}_{g_2}  } \times P_{\rm lin}(k,z) \, \times \\ 
    \int dM\,  n(M,z) \, \langle N_{c,1}(M) \rangle  \Big[ 1 + \,\langle N_{s,1} (M)\rangle\, u(k|M) \Big] \times \\ 
    \int dM\,  n(M,z) \, \langle N_{c,2}(M) \rangle  \Big[ 1 + \,\langle N_{s,2} (M)\rangle\, u(k|M) \Big].
\end{multline}}

We will also compare our fitting results with a 5-parameter HOD (two central and three satellite terms) as introduced in \cite{zheng_galaxy_2007}, using the definition from \cite{nicola_tomographic_2020}. Here we again define the total $\langle N(M) \rangle$ as in Equation \eqref{eq:NM}, but the central distribution is 
\begin{equation}\label{eq:nc5}
    \langle N_c (M)\rangle = \frac{1}{2} \left[ 1- {\rm erf} \left( \frac{\ln(M/M_{\rm min})}{\sigma_{lnM}} \right) \right]
\end{equation}
where $\rm erf$ is the error function ${\rm erf}(x) = \frac{2}{\sqrt{\pi}} \int_0^x e^{-t^2} dt$, and the HOD parameters are: $M_{min}$, which is as previously defined, and $\sigma_{lnM}$ sets the width of the central galaxy profile cutoff. The satellite distribution is defined such that
\begin{equation}\label{eq:ns5}
    \langle N_s (M)\rangle = \Theta(M-M_0) \left( \frac{M-M_0}{M_1}\right)^\alpha,
\end{equation}
where the HOD parameters are: $M_1$, the same as defined for Equation \ref{eq:Ns3}, $M_0$ is the cutoff halo mass below which one does not find satellites, and $\alpha$ is also as previously defined.

From these HOD prescriptions, we must next project the full three-dimensional definition of the galaxy power spectrum to the measured correlation statistic. 

\subsubsection{Projection to the Angular Correlation Function}
We project the galaxy-galaxy power spectrum using the following standard form of Limber's equation for a cross-correlation \citep{limber_analysis_1953, groth_statistical_1977, peacock_power_1991, eisenstein_correlations_2001}:
\begin{align} \label{eq:limber}
\begin{split}
    \omega(\theta) = \pi \int_{z = 0} ^{\infty} \int_{k = 0} ^{\infty} \frac{\Delta^2 (k,z)}{ k^2} J_0[k\, \theta \, \chi(z)] 
    \\
    \times \left(\frac{dN}{dz}\right)_1 \left(\frac{dN}{dz}\right)_2 \left(\frac{dz}{d \chi}\right) dk\, dz, 
\end{split}
\end{align}
where $\Delta^2 (k,z) = \frac{k^3}{2\pi^2} P_{g_1g_2}(k,z)$, and $P_{g_1g_2}(k,z)$ is the galaxy-galaxy power spectrum as detailed in Equation \eqref{eq:Pgg_def}. $J_0$ is the zeroth-order Bessel function, $\chi(z)$ is the comoving distance in units of Mpc, and $dz/d\chi$ is defined using the CCL function with units of length in Mpc. This projection requires the redshift distribution of the galaxy samples, which we define for our weighted clustering approach such that: 
 \begin{equation} \label{eq:wdNdz}
    \frac{dN}{dz} = \sum_i \mathcal{W}_i \cdot p_i(z~|~\mathcal{W}_i \geq \mathcal{W}_0),
\end{equation}
where we have required a lower bound of $p(z)$ and redshift bin overlap of  $\mathcal{W}_0 = 0.03$. Given our wide redshift bin size and survey field area, we will only use the standard Limber approximation, finding that the extended model as defined in \cite{simon_how_2007} is unnecessary. We are thus able to calculate the HOD model's projection for an angular correlation function, and fit for the HOD's parameters using a Bayesian analysis to infer the parameter model posteriors.

\subsubsection{Likelihood and Derived Parameters}\label{sec:fitting}
We compute the standard Gaussian likelihood for our angular correlation function with
\begin{equation}
    \ln \mathcal{L}= -\frac{1}{2} \, (\omega_{\rm data} -  \omega_{\rm model})^T_j \, C_{j,k}^{-1}\, (\omega_{\rm data} -  \omega_{\rm model})_k,
\end{equation}
where $C_{j,k}^{-1}$ is the inverse covariance matrix for the fitted scales of interest, $\omega_{\rm data}$ is the measured angular correlation function and $\omega_{\rm model}$ is the HOD-derived galaxy-galaxy power spectrum projected using Equation \eqref{eq:limber}. We use the MCMC sampler \texttt{emcee} \citep{foreman-mackey_emcee_2013} and will quote parameter uncertainties using the $16^{\rm th}$ and $84^{\rm th}$ percentiles from the posterior of each HOD-estimated and derived parameter. Our principal strategy will be to execute a joint fit of the three correlation functions for a particular galaxy or AGN sample across the three HSC fields, requiring the single best HOD parametrization to fit all the available data. We run these chains until the MCMC-derived integrated autocorrelation time $\tau$ meets the convergence conditions described by \cite{goodman_ensemble_2010} (in brief, the estimated $\tau$ from the chains with $N$ samples crosses the $\tau = N/50$ line). 

We take the HOD parameters from the LRG autocorrelations and fix them when fitting the cross-correlations with the AGN. We impose wide uniform priors on our 3-- and 5--parameter models that are uninformative aside from the upper halo mass limit. These are $\log M_{min} \in [9,\,16.95] \, \log M_\odot$, $\log M_{1} \in [9,\,16.95] \log \, M_\odot$ , $\alpha \in [0,\,4]$, $\log M_{0} \in [9,\,16.95] \,\log M_\odot$, $ \sigma_{\ln M} \in [0,\,4]$, where we fix halo mass upper limit to $\log(M_\odot) < 16.95$ due to numerical limitations in CCL. 

In addition to the fitted HOD parameters, we will also report three derived parameters from the full posterior of the HOD's. These are the satellite fraction:
\begin{equation}\label{eq:fsat}
    f_{sat} = \frac{\int \! \int dz \,\,  dM \, n(M,z)  \, \langle N_c (M)\rangle\,\langle N_s (M)\rangle \, \frac{dN}{dz}}{\int{dz\,\, \bar{n}_g(z) \frac{dN}{dz}}}, 
\end{equation}
which parametrizes the proportion of galaxies in the sample that are satellites of a central galaxy. This definition is specific for the form of the central and satellite profiles as presented in Equation \ref{eq:NM}. We will also estimate the average halo mass, defined such that:
\begin{equation}\label{eq:Mave}
    \langle M_h\rangle  = \frac{\int \!\int dz \,\,  dM\, \, \,  M \,\,  n(M,z)  \,  \langle N (M)\rangle\, \frac{dN}{dz}}{\int{dz\,\, \bar{n}_g(z) \frac{dN}{dz}}}. 
\end{equation}
We examine the separate contributions of the central and satellite galaxies to the total halo mass posterior by splitting the contributions of the one- and two-halo terms. We define $\langle M_{h}\rangle_c$ and $\langle M_{h}\rangle_s$ for the average central and satellite host halo mass by swapping out $\langle N(M)\rangle$ with $\langle N_c(M)\rangle$ and $\langle N_c(M)\rangle \langle N_s(M)\rangle$ into Eq. \eqref{eq:Mave}, respectively. Additionally, we estimate the linear galaxy bias:
\begin{equation}\label{eq:bg}
    b_g = \frac{\int \! \int dz\, \, dM \,\,  b(z,M)  \, \, n(M,z)\, \langle N (M)\rangle \frac{dN}{dz}} {\int{dz \,\, \bar{n}_g(z) \frac{dN}{dz}}},
\end{equation}
to relate our HOD parameters to the linear galaxy bias. 

For purposes of comparing the goodness-of-fit across different HOD parameterizations, and to do so consistently with other AGN angular clustering analyses \citep[c.f.][\citetalias{cordova_rosado_cross-correlation_2024a}]{koutoulidis_clustering_2013, koutoulidis_dependence_2018, dipompeo_angular_2014, dipompeo_updated_2016, dipompeo_characteristic_2017, petter_host_2023}, we define the total $\chi^2$ for the joint-field fit as: 
\begin{equation}
    \chi^2 = \sum_i^3 \, \frac{(\omega_{data} - \omega_{model})_i^2}{\sigma^2_i},
\end{equation}
where $\sigma_i$ is the square root of the diagonal of the covariance matrix for a given field's correlation function. We will also compare fits with the reduced $\chi^2$, such that \begin{equation}
    \chi^2_{\nu}  = \chi^2/N_{\rm dof}.
\end{equation}
For our joint-likelihood fit across HSC fields, we define the degrees of freedom (d.o.f.) as the total number of datapoints minus the number of parameters being fit in the HOD. We caution, however, that $ \chi^2_{\nu}$ is poorly defined for a non-linear model fit \citep[cf. ][]{andrae_dos_2010}.

\section{Results}\label{sec:results}

\begin{figure}
    \centering
    \includegraphics[width=1.\linewidth]{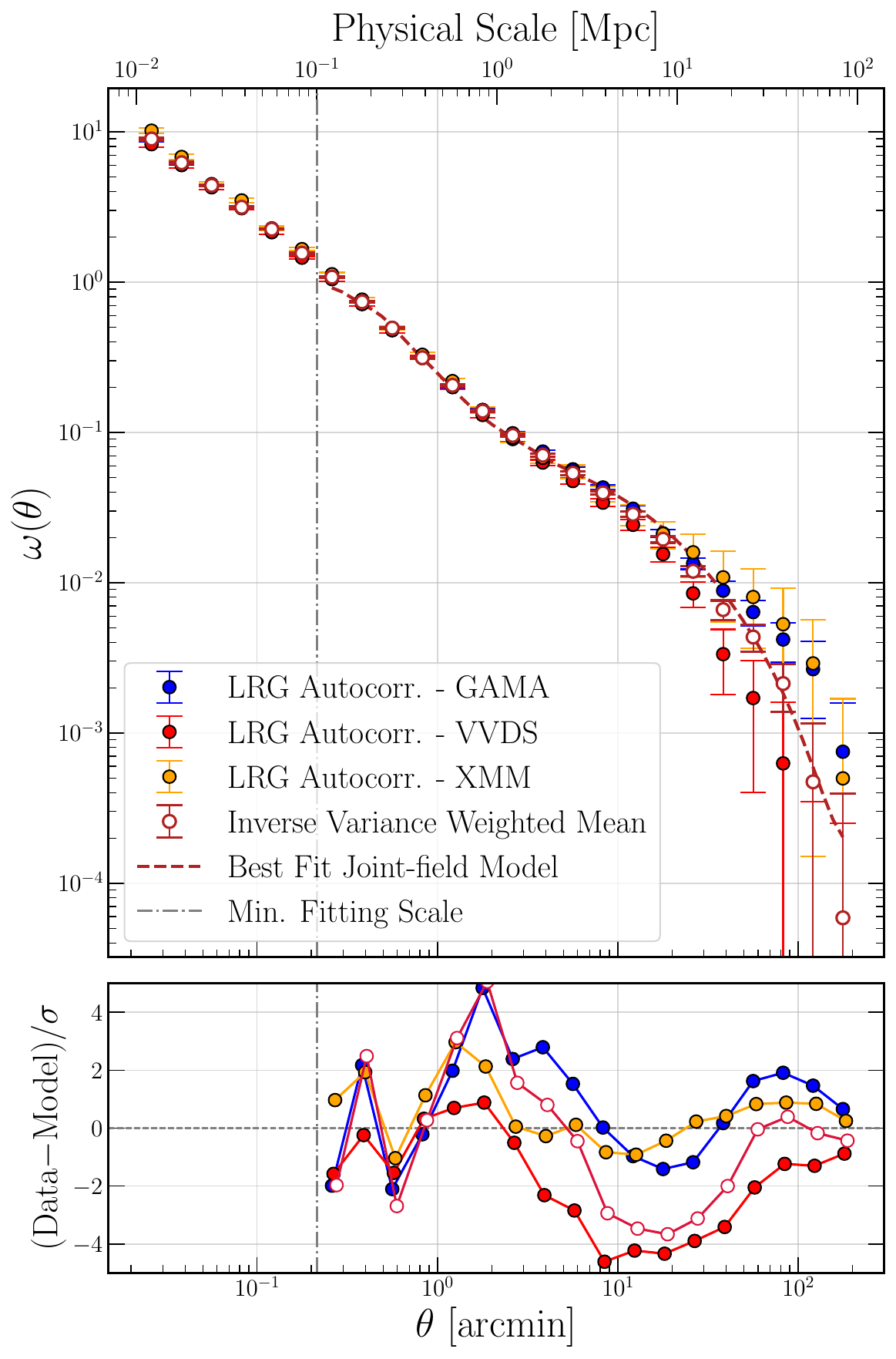}
    \caption{\textit{Top:} The measured HSC LRG projected angular autocorrelation for our three HSC fields. The $1\sigma$ uncertainties are drawn from the square root of the diagonal of the jackknife covariance matrix for each sample. The open symbols represent the per-bin inverse variance weighted mean and error across the fields. The dashed line represents the joint-field best-fit 3-parameter HOD model. The gray dash dotted line represents the minimum scale for which we fit the data, $s > 0.1\,{\rm Mpc}$. \textit{Bottom:} Residuals from each field and their inverse variance-weighted mean, highlighting the poor fit on large scales.}
    \label{fig:summary_ACF_LRG}
\end{figure}

\begin{figure}
    \centering
    \includegraphics[width=1.\linewidth]{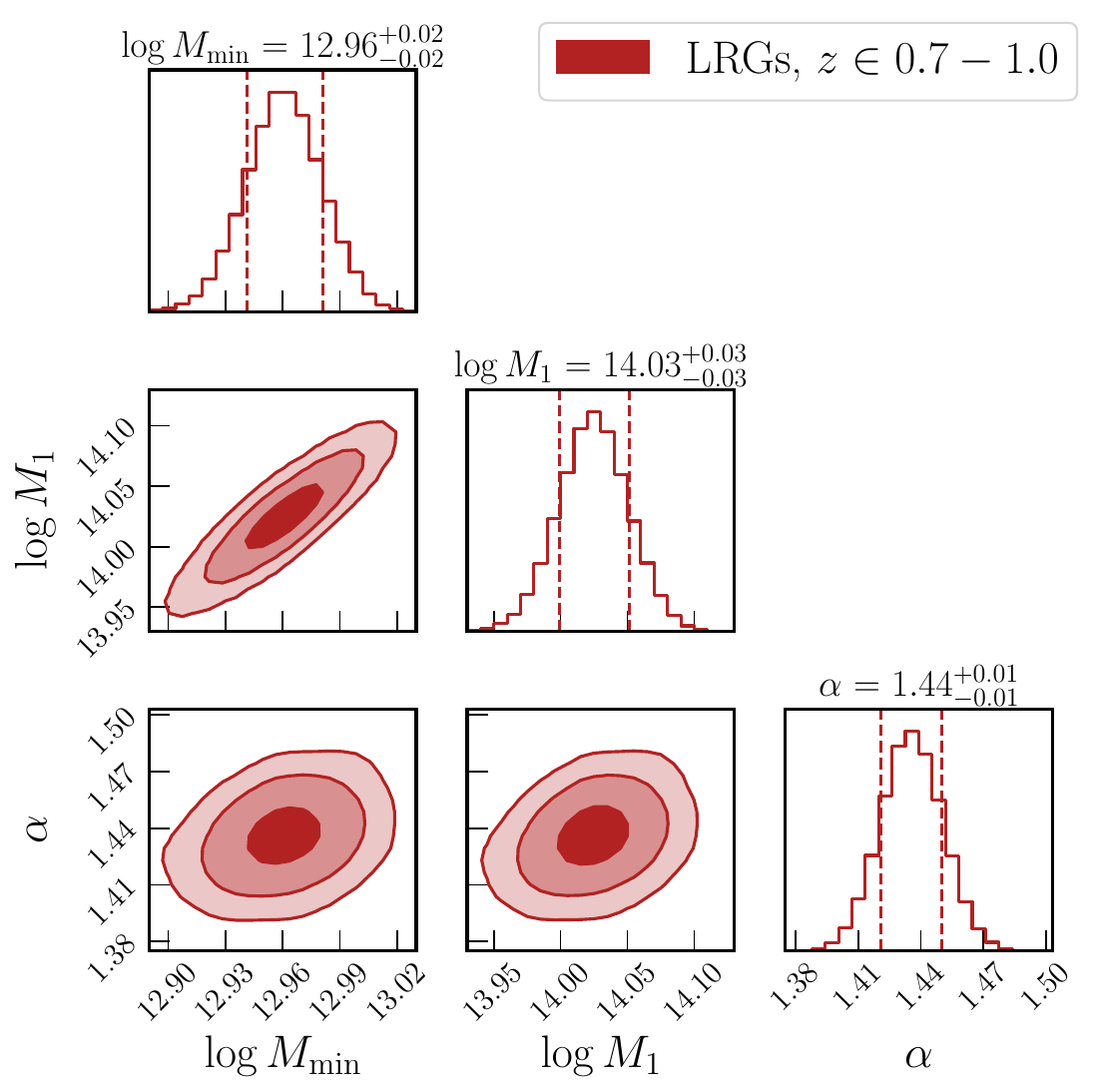}
    \caption{MCMC-derived posteriors for the 3-parameter HOD model fit to the LRG autocorrelations in our $z\in0.7-1.0$ bin (see definitions in \S \ref{sec:HOD_method}). Contours are shown for the 1, 2, and $3\sigma$ 2-D confidence levels ($39.4\%,\, 86.5\%,\,98.9\%$). We maximize the joint likelihood by summing the likelihood of each subfield while requiring that a single HOD model fit all the data. We recover HOD parameters for our magnitude-limited ($r<24$) LRG sample, finding that our derived parameters are $b_g = 2.07 \pm 0.01$, $\log \langle M_h\rangle= 13.48\pm0.01\, \log M_\odot$, and $f_{sat} = 11.1 \pm 0.5\%$ . }
    \label{fig:summary_LRG_corner}
\end{figure}

In this section we discuss the results of fitting the measured angular correlation functions with HOD models. We first fit the LRG autocorrelations. We then use these fitted parameters to fix the LRG component of the cross-correlation HOD and fit only for the AGN component. 

\subsection{HOD Fits for LRG Autocorrelations}\label{sec:LRGauto}
We computed the LRG autocorrelations following the procedures outlined in \S \ref{sec:clustering_methods}, as done previously in \citetalias{cordova_rosado_cross-correlation_2024a}. There, we restricted our analysis to the linear (two-halo term dominated) regime, fitting scales $s\gtrsim1.4\,{\rm Mpc}$. Here, we fit the non-linear (one-halo term dominated) regime to as small a scale as the model is traditionally employed, for $s>0.1 {\rm \, Mpc}$ \citep[c.f.][]{masjedi_very_2006, sawangwit_angular_2011, zhai_clustering_2017}. At scales smaller than this, angular correlation function analyses usually employ a simple power law model to describe the slope of the most interior one-halo term points \citep{masjedi_very_2006, sawangwit_angular_2011}, due to small-scale limitations of HOD modeling. 

The autocorrelations for each HSC field are shown in the top panel of Figure \ref{fig:summary_ACF_LRG}, while the bottom panel shows the residuals from the best-fit (3-parameter) HOD model, with $\mathcal{O}\sim20\% \, (\theta < 10')$ to $\sim50\% \, (\theta > 10')$ variability between fields. In turn, we use the spread of the autocorrelation between fields to estimate systematic uncertainties and cosmic variance between the fields in performing our joint-fit. For illustrative purposes, we plot the per-angular-bin inverse variance weighted mean and error between HSC fields as the open symbols. As we explored in our earlier work (\citetalias{cordova_rosado_cross-correlation_2024a}), the large scale clustering variability is driven by photometric redshift misattributions. We find that the low amplitude of the autocorrelation from the VVDS field is driving the high $\chi^2 = 351$ for 51 d.o.f., and using a 5-parameter HOD model did not improve the joint-field fits. We confirm our prior findings that the variability between fields is a sub-dominant source of error on our final AGN clustering interpretation. The per-field LRG HOD inferred $\omega(\theta)$ differs between HSC fields at the $\leq 10\%$ level, while we note the $\omega(\theta)$ differences between distinct AGN sub-sample HOD models are as much as $\sim60\%$.

Using the formalism described in \S \ref{sec:HOD_method}, we fit the three-parameter HOD with results shown in Figure \ref{fig:summary_LRG_corner}. The estimated HOD parameters are: $\log\, M_{\rm min} = 12.96\pm 0.02 \, \log M_\odot$, $\log\, M_{1} = 14.03\pm 0.03 \, \log M_\odot$, and $\alpha = 1.43 \pm 0.01$. Taking the posteriors for each HOD parameter, we calculate the derived parameters using Equations \eqref{eq:fsat}, \eqref{eq:Mave} and \eqref{eq:bg}, finding $b_g = 2.07 \pm 0.01$, $\log \langle M_h\rangle= 13.48\pm0.01\, \log M_\odot$, and $f_{sat} = 11.1 \pm 0.5\%$ for our LRG population. We find consistent HOD parameters with other LRG clustering analyses in this redshift range \citep[$\langle M_h\rangle \approx 10^{13.5} M_\odot$][]{sawangwit_angular_2011, ishikawa_halo-model_2021, zhou_clustering_2021}, though direct comparisons are difficult given that our sample is defined in terms of flux, rather than stellar mass, thresholds. The inferred properties of these correlation functions are summarized in the leftmost column of Table \ref{tab:HOD_results}. Having solved for the LRG HOD, we now consider the HOD fits for the cross-correlation between LRGs and AGN.

\subsection{HOD Fits for Full AGN Sample Cross-correlations}

\renewcommand{\arraystretch}{1.3}

\begin{table*}
\centering
\caption{Derived HOD Parameters for HSC+$WISE$ AGN}
\begin{tabular}{c|c|ccccccc}
\hline \hline
& & & \multicolumn{6}{c}{AGN Samples} \\
{Parameter} & {LRGs} & {All} & {Unobsc.} & {Unobsc.+Redd.} & {Redd.} & {Redd.+Obsc.}& {Obsc.} \\

\hline
$N_{obj}$ & 1,509,905 & 28,494  & 8,266 & 15,156 & 6,890 & 19,675 & 12,785 \\

Weighted $N_{obj}$ & 843,166.6  & 13,898.8    & 3,942.0 & 6,745.71& 2,803.6 & 9,893.4 & 7,089.8 \\

{$\langle z \rangle$} & $0.8\pm0.1$  & $0.9^{+0.1}_{-0.2}$  & $0.9^{+0.1}_{-0.2}$ & $0.9^{+0.1}_{-0.2}$ & $0.9^{+0.1}_{-0.2}$ & $0.9^{+0.1}_{-0.2}$ & $0.9\pm{0.2}$ & \\

{$\chi^2$ [51 d.o.f.]} & 351  &  75.8  & 60.9 &  69.9 & 46.6 & 60.5 & 54.7 & \\

\hline
$ \log M_{\rm min} $[$\log M_\odot$]& $12.96\pm{0.02}$  &  $12.5 \pm 0.2$ & $13.21^{+0.04}_{-0.06}$  & $12.9\pm0.2$  & $12.4^{+0.3}_{-0.4}$ &$12.0\pm0.3$ & $12.0\pm0.4$ \\

$ \log M_{1} $[$\log M_\odot$]& $14.02\pm0.03$  &  $13.7^{+0.2}_{-0.1}$ &  $15.5^{+1.0}_{-0.8}$ &  $14.1\pm0.2$ &  $13.6\pm0.3$&  $13.2\pm0.3$ &  $13.0^{+0.5}_{-1.0}$  \\

$ \alpha $ & $1.43\pm0.01$  &$1.2\pm0.2$ &$2.4\pm1.1$ &$1.5\pm0.3$ &$1.5^{+0.3}_{-0.4}$ &$1.1^{+0.2}_{-0.4}$ &$0.7\pm0.5$   \\
\hline
$ b_g $& $2.07\pm0.01$ & $1.7\pm0.1$ & $2.25^{+0.04}_{-0.06}$ & $2.0\pm0.1$ & $1.8\pm0.2$ & $1.5\pm0.1$ & $1.4\pm0.1$   \\

$ \log \langle M_h \rangle\, $[$\log M_\odot$]& $13.48\pm{0.01}$  &  $13.17\pm0.03$ & $13.43^{+0.02}_{-0.03}$  & $13.35\pm0.04$  & $13.27\pm0.05$ &$13.03^{+0.04}_{-0.05}$& $12.91^{+0.07}_{-0.08}$ \\

$ \log \langle M_{h} \rangle_c\, $[$\log M_\odot$]& $13.29^{+0.01}_{-0.02}$  &  $12.9\pm0.1$ & $13.43^{+0.03}_{-0.04}$  & $13.2\pm0.1$  & $12.9^{+0.2}_{-0.3}$ &$12.6\pm0.2$& $12.6^{+0.2}_{-0.3}$ \\

$ \log \langle M_{h} \rangle_s\, $[$\log M_\odot$]& $14.07\pm0.01$  &  $13.8\pm0.01$ & $14.4^{+0.3}_{-0.4}$  & $14.0^{+0.2}_{-0.1}$  & $13.9\pm0.2$ &$13.6^{+0.2}_{-0.3}$& $13.2^{+0.4}_{-0.3}$ \\

$ f_{sat} $ & $11.1 \pm 0.5\%$ & $13^{+3}_{-2} \%$ & $0.05^{+1}_{-0.05} \%$ & $8\pm3\%$ & $15^{+6}_{-4}\%$ & $20^{+10}_{-5}\%$ & $31^{+23}_{-14}\%$\\

\hline \hline
\end{tabular}\label{tab:HOD_results}
\vspace{3mm}

\end{table*}

\begin{figure}[t]
    \centering
    \includegraphics[width=1.\linewidth]{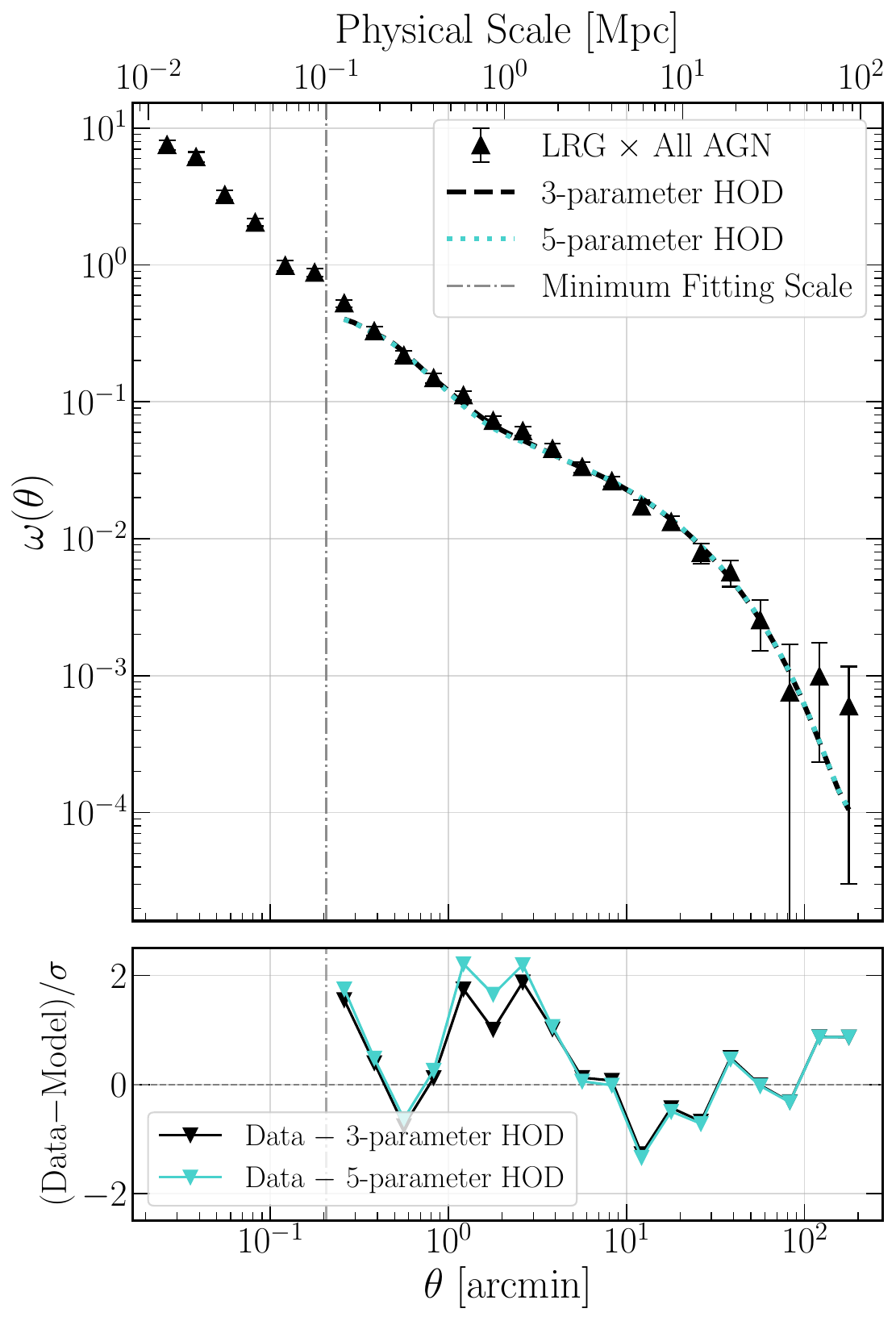}
    \caption{\textit{Top:} The inverse-variance weighted mean of the measured LRG $\times$ full HSC+\textit{WISE} AGN sample projected angular autocorrelation. We co-add the three HSC fields' measured cross-correlations and illustrate them as black triangles. The $1\sigma$ uncertainties are the inverse variance weighted error. The black dashed line represents the joint-field best-fit 3-parameter HOD model, while the dotted turquoise line is the same for the 5-parameter HOD model. The gray dash dotted line represents the minimum scale for which we fit the data, $s > 0.1\,{\rm Mpc}$. \textit{Bottom:} Residual plots for the 3- and 5-parameter best-fit HOD models (colors as in the above panel). }
    \label{fig:All_AGN_HOD_comparison}
\end{figure}

\begin{figure}
    \centering
    \includegraphics[width=1.\linewidth]{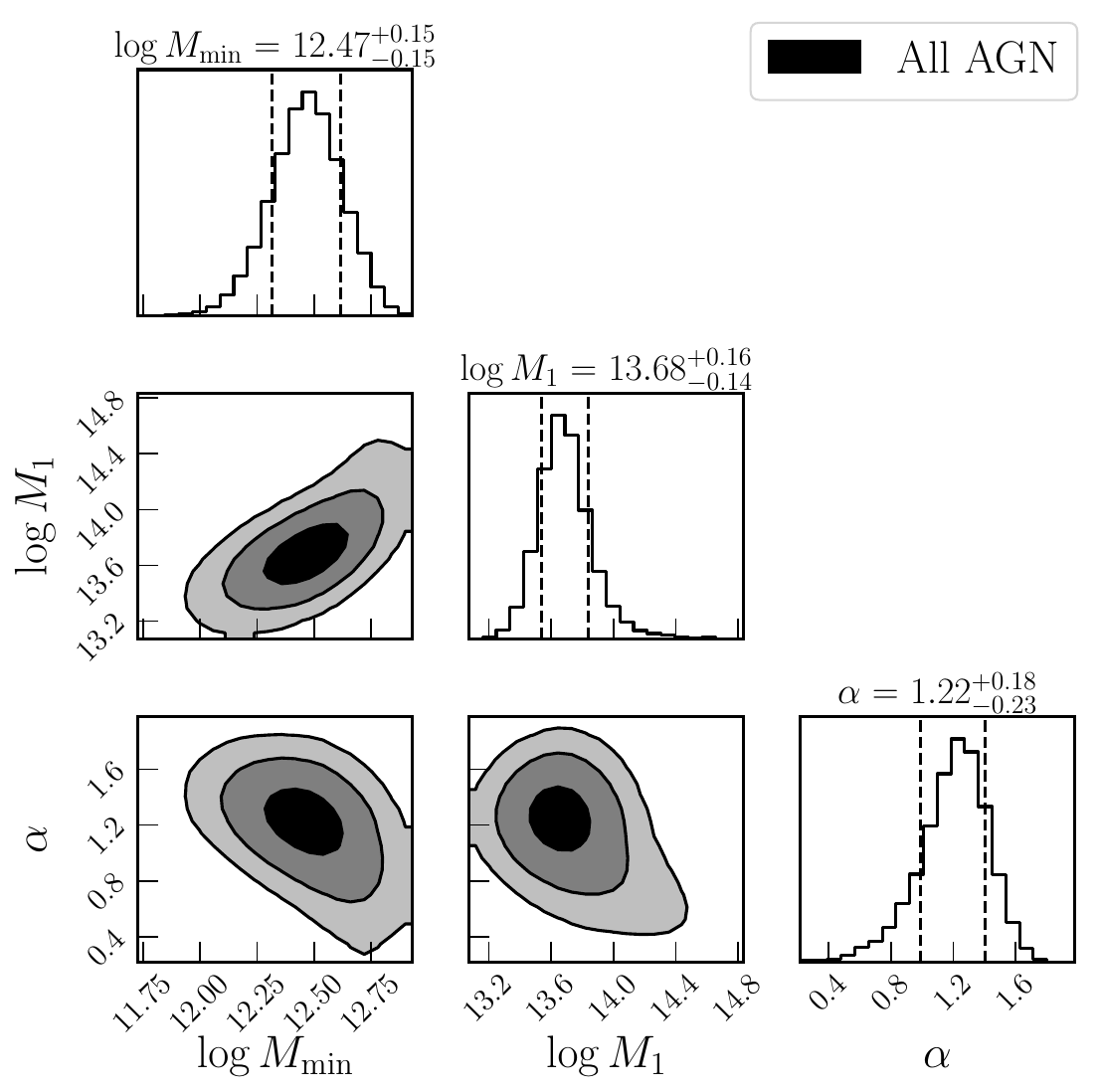}
    \caption{Posteriors for the 3-parameter HOD model fit to the cross-correlation of LRGs and the complete AGN sample in our $z\in0.7-1.0$ bin. Contours are shown for the 1, 2, and $3\sigma$ 2-D confidence levels ($39.4\%,\, 86.5\%,\,98.9\%$). After fixing the cross terms for the LRGs from our fits to their autocorrelation, we maximize the joint likelihood by summing the likelihood of each subfield while requiring a single AGN HOD model fit all the data. Our derived parameters are $b_g = 1.7 \pm 0.1$, $\log \langle M_h\rangle= 13.17\pm0.03\, \log M_\odot$, and $f_{sat} = 13^{+3}_{-2} \% $. }
    \label{fig:summary_AGN_All_corner}
\end{figure}

\begin{figure*}[t!]
    \centering
    \includegraphics[width=1.\linewidth]{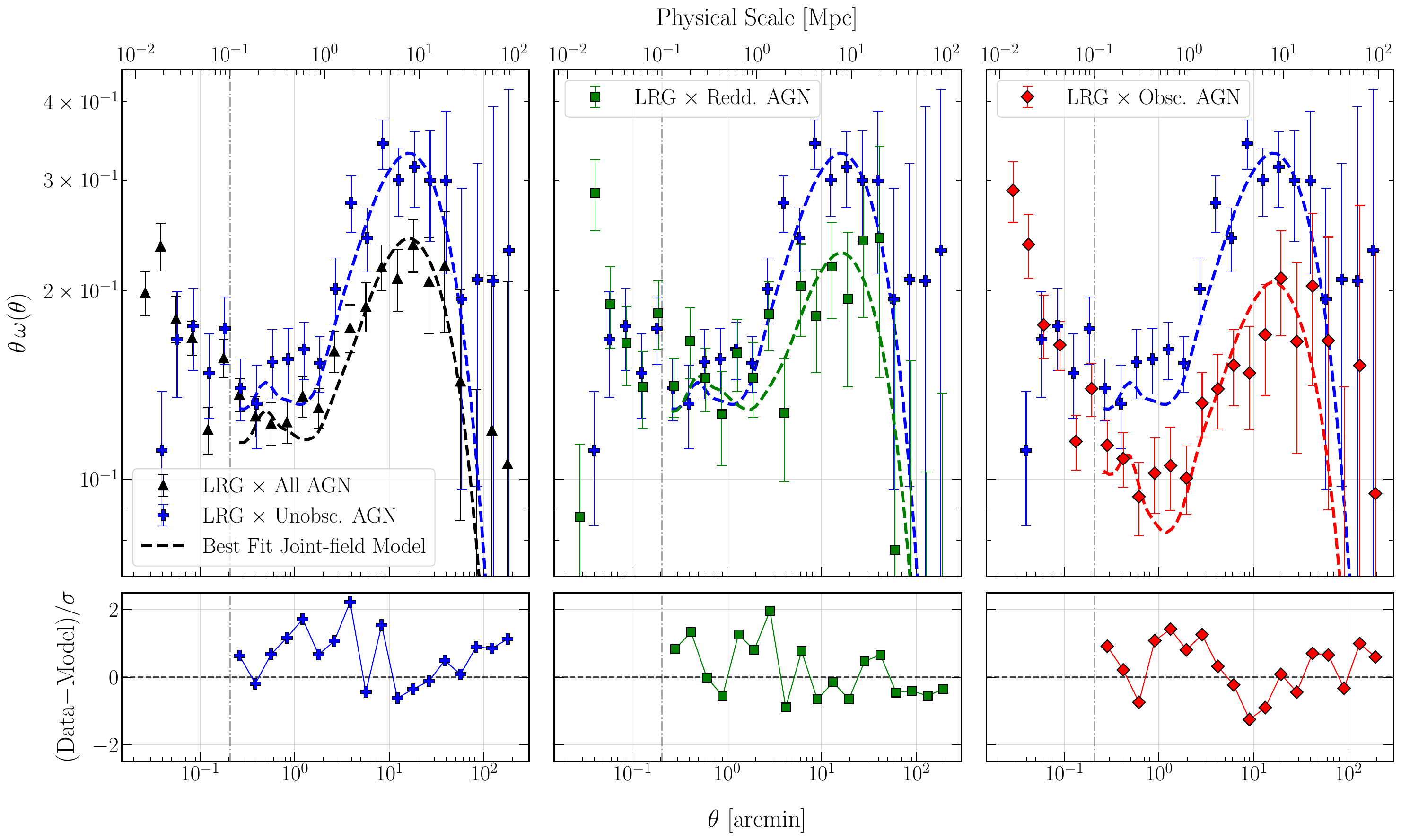}
    \caption{\textit{Top:} Cross-correlation between HSC LRGs and the distinct AGN sub-type samples, each fitted with a 3-parameter HOD, shown as $\theta\, \omega(\theta)$ to reduce the plotted dynamic range. The AGN sample is limited to $L_{6\mu m} > 3\times 10^{44} \, {\rm erg\, s^{-1}}$. We show the inverse-variance-weighted mean and error for each sample across the three fields; the dashed line represents the best fit 3-parameter HOD to the data. We split the subsets among the panels, keeping the cross-correlation of LRGs and unobscured AGN throughout as a comparison. The vertical gray dash-dotted line represents the minimum scale for which we fit the data, $s > 0.1\,{\rm Mpc}$. \textit{Bottom:} Residual plots for the fitted models to each of the AGN sub-samples.}
    \label{fig:summary_ACF}
\end{figure*}

\begin{figure}
    \centering
    \includegraphics[width=1.\linewidth]{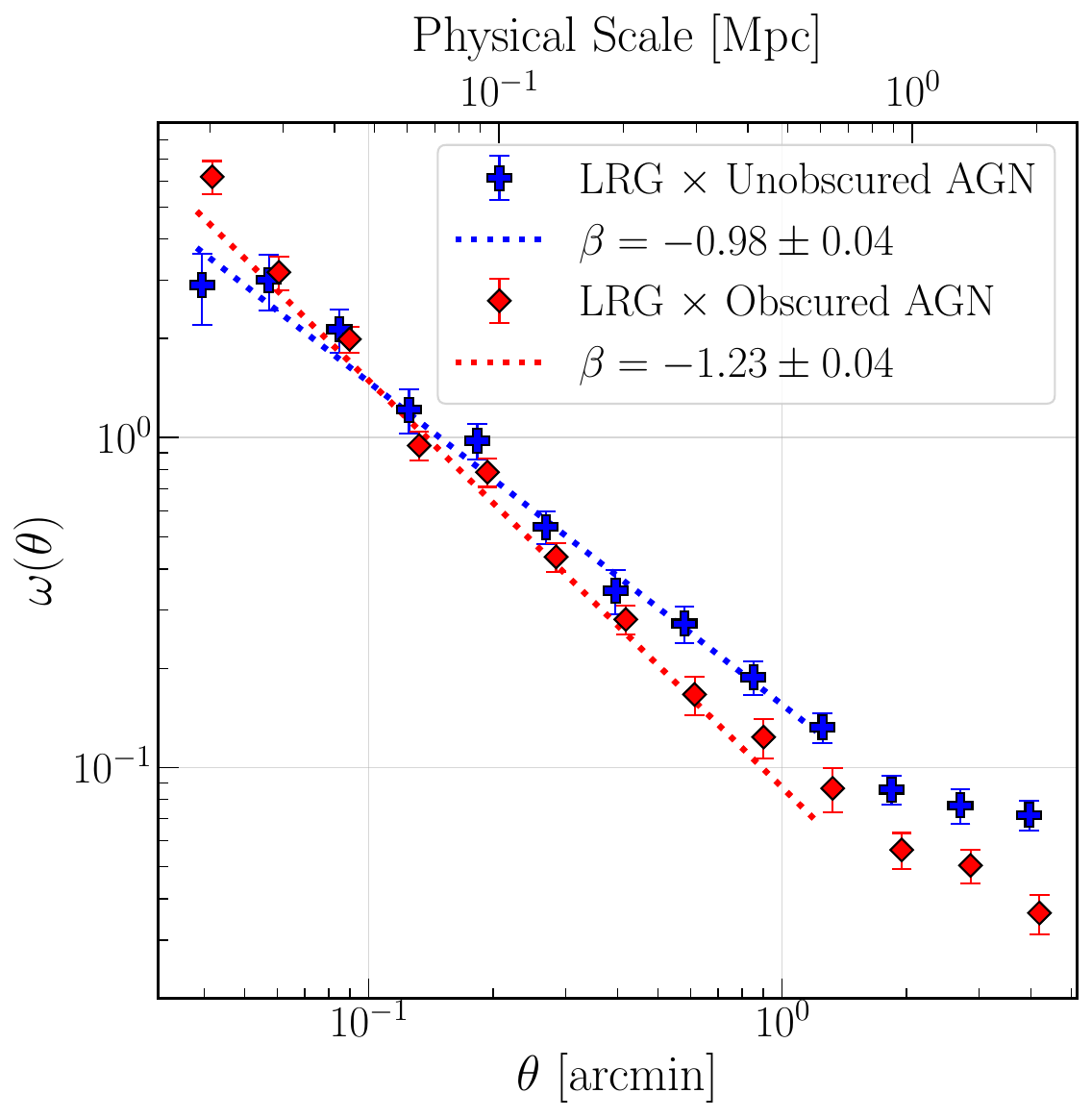}
    \caption{Detailed view of the one-halo angular regime of the inverse variance-weighted mean angular correlation function for unobscured and obscured AGN in cross-correlation with LRGs. We fit a simple power law model on scales where the one-halo term dominates, $0.015<s<0.7 \, {\rm Mpc}$, and the fits are shown as dotted lines. The best-fit slope $\beta$ of one correlation function rejects the other at $>6 \sigma$. The slope of the obscured (Type II) AGN is steeper than that of unobscured (Type I) AGN, suggesting objects in our Type II sample populate satellite in halos more frequently.}
    \label{fig:power-law-fit}
\end{figure}

Using the complete sample of $L_{6\mu m}$ luminosity-limited HSC+\textit{WISE} AGN for $z\in0.7-1.0$ we described in \S \ref{sec:dat_AGN}, we presented the cross-correlation between these and our LRG sample in \citetalias{cordova_rosado_cross-correlation_2024a}. Taking the measurements from projected physical scales $s>0.1 {\rm \, Mpc}$, we again follow the modeling and fitting procedure outlined in \S \ref{sec:HOD_method} to jointly fit the HOD for the full AGN sample across our three HSC fields. To more cleanly illustrate the overall clustering, we plot the inverse variance weighted mean and error of the per-field LRG $\times$ All AGN cross-correlations in the top panel of Figure \ref{fig:All_AGN_HOD_comparison}. 

For these cross-correlations, we fix the parameters of the LRG component of the HOD model from our fits to their autocorrelations (see \S \ref{sec:LRGauto}). We use the median value for each LRG HOD parameter posterior, finding that the LRG parameters' uncertainty contributes $<2\%$ of the error for our AGN HOD parameters. We emphasize that the field-to-field systematic uncertainties from the LRG autocorrelation are sub-dominant to the uncertainties found from our AGN cross-correlations. Our ultimate goal of investigating if there are any relative differences between AGN sub-samples' clustering is less sensitive to the LRG catalog systematics given we fix the LRG HOD parameters consistently across AGN sub-types. 

Next, we compare the fits from the 3-parameter and 5-parameter HOD models to our measured correlation function. We plot the derived best-fit models for each parameterization as black dashed (3-parameter HOD) and turquoise dotted lines (5-parameter HOD), observing that both models effectively fit the intra-halo clustering. We find that the 3-parameter HOD has a $\chi^2_\nu = 75.8/51 \approx 1.5$. Using the 5-parameter HOD, we recover a $\chi^2_\nu = 83.2/49 \approx 1.7$. Based on these values, we perform a Gaussian likelihood ratio test and find the models have a $\Delta\sigma \approx 0.4$ preference for the 3-parameter model. We will use the 3-parameter HOD model for the remainder of this analysis as our default, finding that the fit does not improve with the inclusion of additional parameters. We emphasize the subtle differences between these best-fit HOD models in the bottom panel of Figure \ref{fig:All_AGN_HOD_comparison}. Comparable angular correlation function studies \citep[cf. ][]{sawangwit_angular_2011} have similarly not found a statistically-motivated reason to use a 5-parameter model when analyzing $\omega(\theta)$ measurements.

We show the posteriors for the HOD fits for the complete AGN sample in Figure \ref{fig:summary_AGN_All_corner}, and include their median and $16^{\rm th}$ and $84^{\rm th}$ percentile errors in Table \ref{tab:HOD_results}. The estimated HOD parameters are: $\log\, M_{\rm min} = 12.5\pm 0.2 \, \log M_\odot$, $\log\, M_{1} = 13.7^{+0.2}_{-0.1} \, \log M_\odot$, and $\alpha = 1.2\pm0.2$. Using these posteriors we calculate the derived parameters and find $b_g = 1.7 \pm 0.1$, $\log \langle M_h\rangle  = 13.17\pm0.03 \, \log M_\odot$, and $f_{sat} = 13^{+3}_{-2}\%$. This inferred galaxy bias from the HOD parameter fit is consistent with the linear bias fits we showed in \citetalias{cordova_rosado_cross-correlation_2024a} (linear clustering bias for ``All AGN'' sample, $b_g = 1.4\pm0.2$). The derived satellite fraction for our complete AGN sample is approximately $35\%$ lower than estimates of the LRG $f_{sat}$ for a similar redshift range and $M_{min}$ \citep[cf. Table 2 of ][]{ishikawa_halo-model_2021}, though only at $\sim2\sigma$ statistical significance. We next estimate HOD parameters for the different AGN sub-type samples.

\subsection{HOD Fits for Cross-correlations of AGN Subtypes}

\begin{figure*}[t!]
    \centering
    \includegraphics[width=1.\linewidth]{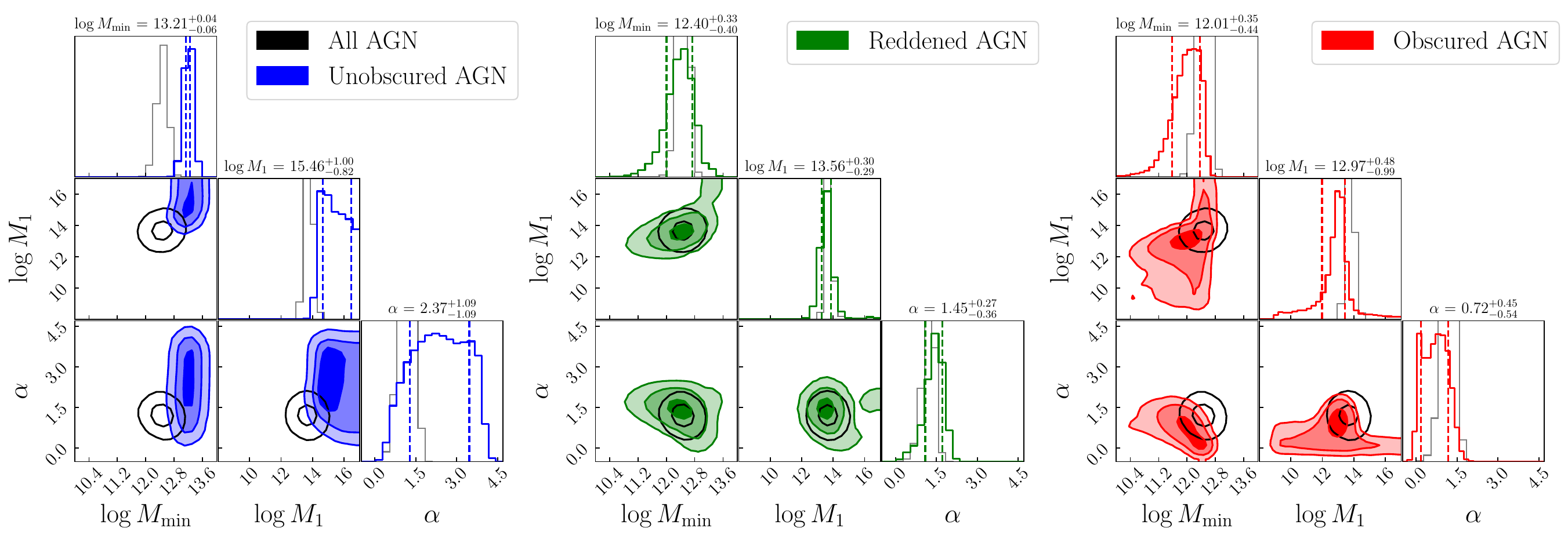}
    \caption{MCMC-derived posteriors for the 3-parameter HOD model (defined in \S \ref{sec:HOD_method}) fit to cross-correlations of LRGs with the unobscured (\textit{left}), reddened (\textit{middle}), and obscured (\textit{right}) AGN samples. Contours are shown for the 1, 2, and $3\sigma$ 2-D confidence levels ($39.4\%,\, 86.5\%,\,98.9\%$). The samples are defined for our $z\in0.7-1.0$ bin, and $L_{6\mu m } > 3 \times 10^{44} \, {\rm erg \, s^{-1} }$ threshold. We fix the LRG terms of the cross-correlation HOD model from the above-derived HOD parameters, and maximize the joint likelihood by summing the likelihood of each subfield while requiring a single HOD model fit all the data. The 1 and 2$\sigma$ contours for each parameter from the full AGN HOD parameter estimation are reproduced across the three panels to contextualize the results. Parameter values and 1$\sigma$ uncertainties for each subsample are quoted above the 1-D histograms, and their derived parameters are quoted in Table \ref{tab:HOD_results}. The unobscured and obscured samples' recovered parameter space overlap minimally.}
    \label{fig:corner_subtypes}
\end{figure*}

\begin{figure}
    \centering
    \includegraphics[width=1.\linewidth]{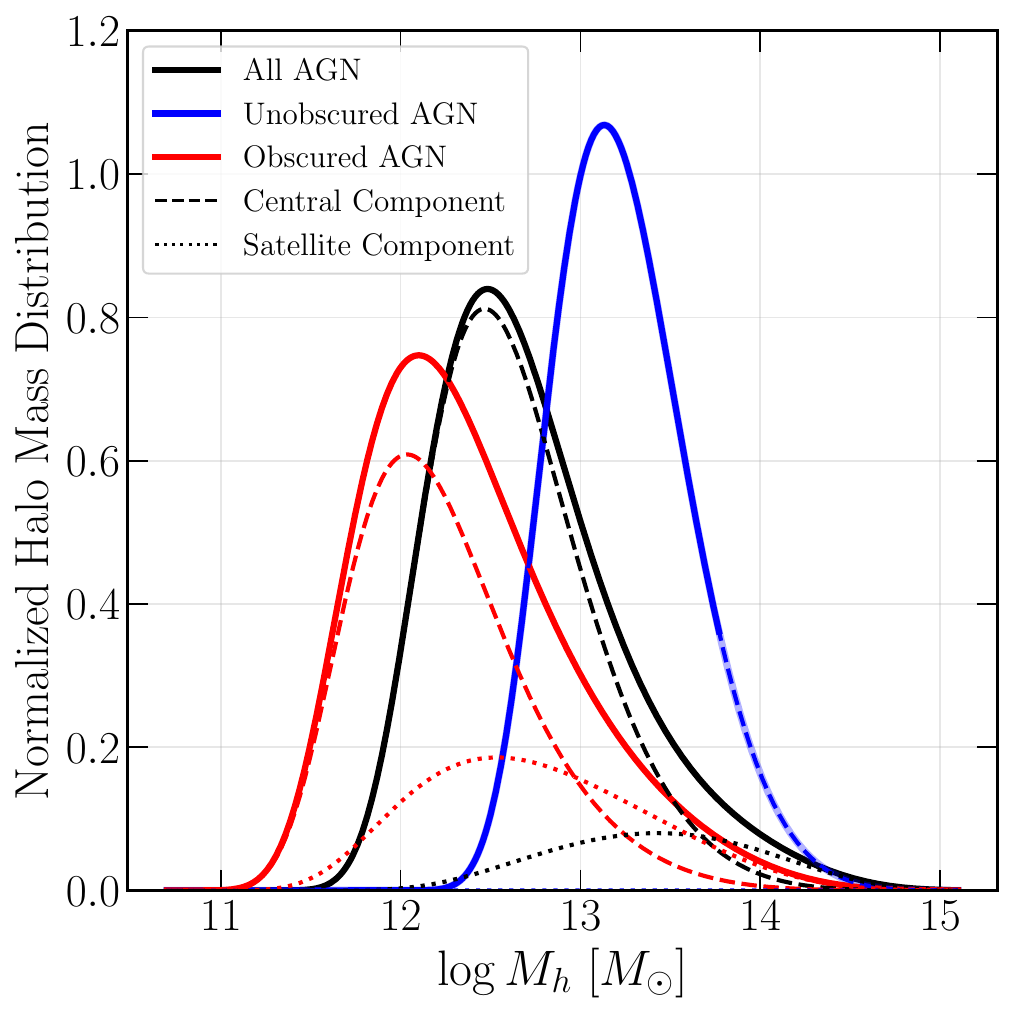}
    \caption{Normalized halo mass distributions for our unobscured, obscured, and all AGN samples. They are calculated by multiplying the \cite{tinker_large-scale_2010} halo mass function, $n(M)$, and the AGN subtype-specific $\langle N(M)\rangle$, divided by $\bar{n}_g$. We define $\langle N(M)\rangle$ with the median HOD parameters from each AGN sub-sample's estimated posteriors. We show the full halo mass distributions (solid lines) as well as the central (dashed line) and satellite galaxy (dotted line) component, for the complete AGN, unobscured AGN, and obscured AGN samples. Note that the unobscured AGN have no significant satellite galaxy contribution to their halo mass distribution.}
    \label{fig:halodists}
\end{figure}

Following full AGN sample's cross-correlation with the LRGs, we fix the joint-field best fit LRG HOD parameters to solve for the HOD of our unobscured, reddened, and obscured AGN samples. These are the same LRG cross-correlations with the $L_{6\mu m} > 3\times 10^{44} \, {\rm erg \,s^{-1}}$ AGN samples presented in \citetalias{cordova_rosado_cross-correlation_2024a}. We investigate if there are any differences in the best-fit HOD and derived parameters between the distinct (photometrically-classified) AGN sub-type samples. We present the angular cross-correlations for all, unobscured, reddened, and obscured AGN with LRGs in Figure \ref{fig:summary_ACF}, showing the inverse variance-weighted mean and error across the HSC fields. The $\omega(\theta)$ values have been scaled by a power of $\theta$ to improve legibility, and the unobscured AGN sample's measurements are repeated across the three panels to serve as a direct comparison. 

The amplitude of the two-halo term ($s\gtrsim1 \, {\rm Mpc}$) is substantially different for each subsample. As we discussed in \citetalias{cordova_rosado_cross-correlation_2024a}, the unobscured AGN cross-correlations have the largest amplitude at linear scales, while obscured AGN have the smallest amplitude. In this analysis, we endeavor to understand each sample's one-halo term properties. We note how the small-scale correlation of the LRG $\times$ unobscured AGN cross-correlation flattens at small scales relative to the other independent AGN samples. To quantify the observed difference, we fit a simple power law to the inverse-variance weighted mean and error from our correlation functions. We consider angular scales dominated by the one-halo term ($0.015<s<0.7 \, {\rm Mpc}$) to estimate the slope, shown in Figure \ref{fig:power-law-fit} as a dotted line. The LRG $\times$ unobscured AGN $\omega^{1h}(\theta)$ has a slope of $\beta = -0.98\pm0.04$, while the LRG $\times$ obscured AGN correlation function has a slope of  $\beta = -1.23\pm0.04$. This difference persists, at a $2\sigma$ significance, when we restrict the fit to ($0.1<s<0.7 \, {\rm Mpc}$). We now turn to DM modeling analyses to connect these empirical differences to physical quantities. 

We can quantify the cross-correlations and their differences as a function of the AGN spectral type with the HOD. The best-fit 3-parameter HOD models are shown as dashed lines in Figure \ref{fig:summary_ACF}. As before, we only fit over the standard HOD scales, at $s>0.1 \, {\rm Mpc}$, indicated by the vertical dotted dash line in Figure \ref{fig:summary_ACF}. The posteriors for each of the HOD fits to the cross-correlations are shown in Figure \ref{fig:corner_subtypes}. We overlay the full AGN sample's $2\sigma$ posteriors across the three sub-figures to contextualize each cross-correlation's best-fit parameters. We record the median and errors for each of our estimated and derived parameters in Table \ref{tab:HOD_results}. For each of our sub-samples, including the linear combinations such as the Unobsc. $+$ Redd and Redd. $+$ Obsc., we find that the inferred $b_g$ values from the HOD are consistent with the directly fit values in our earlier work fitting solely the two-halo term (\citetalias{cordova_rosado_cross-correlation_2024a}). The normalized halo mass distributions for the different AGN sub-samples are shown in Figure \ref{fig:halodists}. We also present their central and satellite contributions, illustrating the lack of a satellite component for the unobscured AGN. 

We compare parameters that summarize the inferred one-halo and two-halo properties. Figure \ref{fig:lit_fsat_comp} shows the posteriors for $\log \langle M_h \rangle$ and $f_{sat}$ for our AGN sub-samples. The inferred average halo mass hierarchy as a function of AGN spectral type we described in \citetalias{cordova_rosado_cross-correlation_2024a, cordova_rosado_cross-correlation_2024b} is again constrained to high significance with this fitting method. We find that $\langle M_h \rangle$ for the unobscured AGN is $\sim3\times$ larger than that for the obscured AGN sample, a $>5\sigma$ difference, while it is $\sim1.5\times$ larger than the reddened AGN, at $3\sigma$ significance. However, when we solely compare the $\langle M_h \rangle_c$ for these samples, the median central galaxy contribution to the halo mass distribution from unobscured AGN is $\sim 6 \times$ more than that for obscured AGN, at $4\sigma$. 

The $f_{sat}$ parameter is determined by the amplitude of the one-halo term from the AGN samples, defined as the fraction of galaxies in the sample which are satellites of a central galaxy (see Equation \ref{eq:fsat}). We find that the unobscured AGN sample has a median $f_{sat} = 0.05^{+1}_{-0.05}\%$, with a mode of $\approx 1\%$ for its skewed distribution. This result suggests that Type I AGN are rarely present as satellites to a central galaxy. Meanwhile, reddened and obscured AGN show more typical values, with median satellite fractions of $15^{+6}_{-4}\%$ and $31^{+23}_{-14}\%$, respectively \citep[cf.][]{zehavi_galaxy_2011, krolewski_tomographic_2025}. We find that there is a $>3\sigma$ difference between the $f_{sat}$ distributions between the unobscured and reddened samples, and a $>3\sigma$ difference in the satellite fraction between the unobscured and the redd. + obsc. samples ($f_{sat} = 20^{+10}_{-5}\%$).

We consider several avenues to verify that our results are not the result of systematic uncertainties, unwise modeling choices, or sample contamination in Appendix \ref{sec:robustness}. It includes tests of alternate HOD formalisms (with 5- and 8- parameter models), the preference for low $f_{sat}$ in Type I AGN, photometric redshift uncertainties, whether this result is confined to our high-$L_{6\mu m}$ sample, and estimates of our possible contamination fraction. We find that our results are robust to all the explored possible sample and analysis choices. We conclude that our measured AGN correlation functions are representative of the underlying clustering for luminosity-limited optical/MIR-selected and classified AGN.

\section{Discussion}\label{sec:disc}

\begin{figure*}
    \centering
    \includegraphics[width=1.\linewidth]{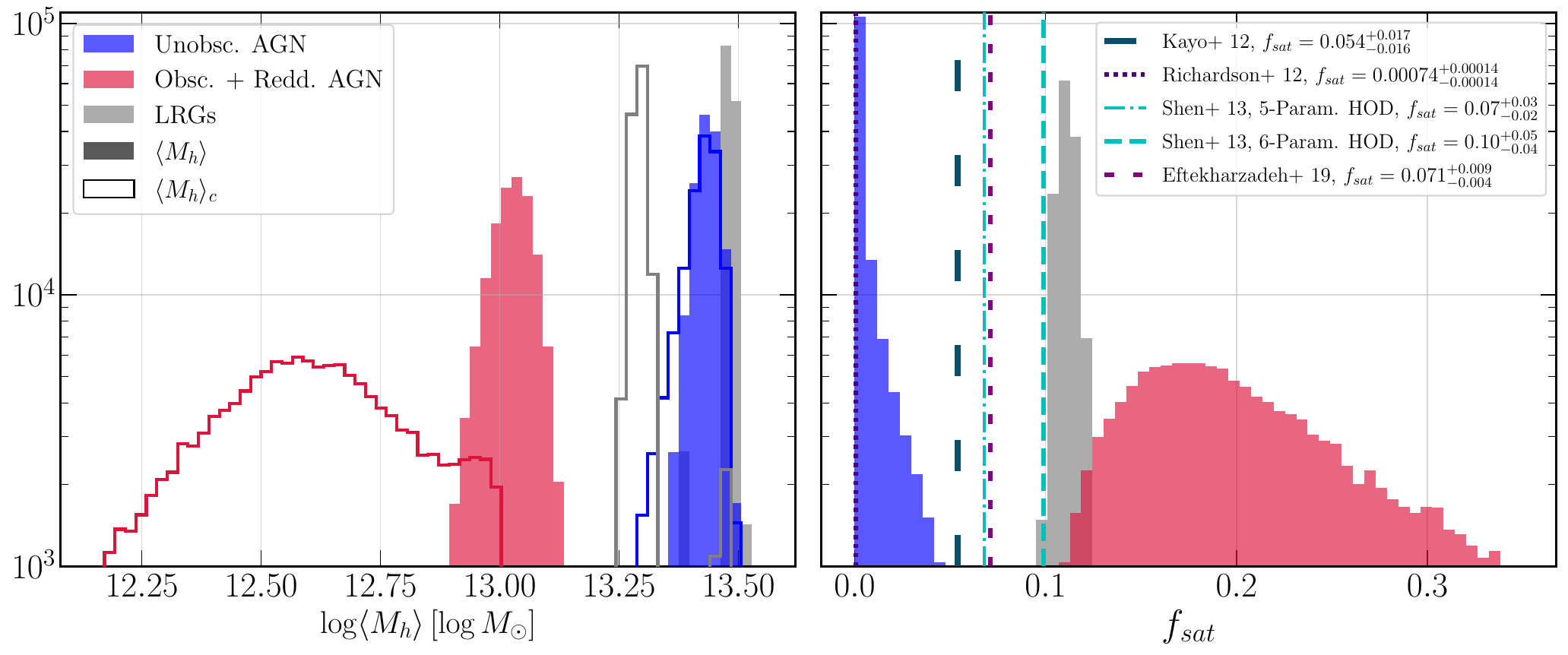}
    \caption{Posterior distributions of derived HOD parameters. The median and uncertainties of these distributions are detailed in Table \ref{tab:HOD_results}. \textit{Left:} The average halo mass $\langle M_h \rangle$ (shaded bars), and central galaxy average host halo mass $\langle M_h \rangle_c$ distributions (solid outlines) for the unobscured, obscured $+$ reddened, and LRG best-fit HODs. We additionally split the derived posteriors for the average total, $\langle M_h\rangle$ (filled bars), and central, $\langle M_h\rangle_c$ (outlined bars), halo mass for each of our galaxy samples. As we found previously in \citetalias{cordova_rosado_cross-correlation_2024a}, our sample of unobscured AGN has a characteristic central galaxy host halo mass that is considerably ($\sim 6\times$) larger than the average AGN with significant obscuration, with a $4\sigma$ statistical significance, while the average halo mass for unobscured AGN is $\sim 3\times$ larger than that of obscured + reddened AGN, at $>5\sigma$ significance. \textit{Right:} The inferred satellite fraction $f_{sat}$ posterior distributions for the unobscured, obscured $+$ reddened, and LRG best-fit HODs. We also show representative values from comparable analyses of the AGN one-halo term. Our analysis is the first to show an HOD-derived $f_{sat}$ for Type II AGN. We note how previously measured Type I $f_{sat}$ values consistently fall below the distribution of our obscured $+$ reddened AGN, as well as below our LRG sample's satellite fraction distribution. The values and uncertainties of the comparable Type I $f_{sat}$ are detailed in the legend.}
    \label{fig:lit_fsat_comp}
\end{figure*}

This investigation effectively measures the one- and two-halo properties of HSC+\textit{WISE} AGN in cross-correlation with luminous red galaxies. The results build on our earlier analyses, and extend our understanding of the physical halo environment in which AGN are triggered. As in our previous analysis of these angular correlation functions, we calculate $\omega(\theta)$ for a single $z\in 0.7-1.0$ redshift bin, and control for the sample completeness by establishing a $L_{6 \mu m} > 3\times 10^{44} \,{\rm erg \,s^{-1}}$ threshold across all the samples, as shown in \S 2.4 of \citetalias{cordova_rosado_cross-correlation_2024a}. We find that the amplitude of the two-halo clustering for Type I AGN is substantially larger ($b_g = 2.25^{+0.04}_{-0.06}$) than for Type II AGN ($b_g = 1.4\pm0.1$), while the Type I one-halo term has a significantly shallower slope ($\beta = -0.98\pm 0.04$) than that of Type II's ($\beta = -1.23\pm 0.04$). We interpret these clustering function shapes with an HOD (see Table \ref{tab:HOD_results}). We show the posteriors on the derived parameters $\langle M_h \rangle$ and $f_{sat}$ in Figure \ref{fig:lit_fsat_comp}.

\subsection{Luminous Type I and II AGN Have Different Clustering Statistics}\label{sec:disc_12}

With high number density catalogs ($\sim 60 \, {\rm AGN \, deg^{-2}}$ in total for our redshift bin), accurate redshift estimation, and low sample contamination rates, we now turn to interpretation of our results. From our collection of systematics tests (see \S \ref{sec:robustness}), we conclude that the correlation functions and HOD fit results that we measure are accurate for optical/MIR-selected, and high $L_{6\mu m}$-limited, unobscured, reddened, and obscured active galactic nuclei. 

We measure and fit the properties of the two-halo term in AGN clustering using our HOD, inferring that the unobscured AGN are hosted, on average,  in $\sim 3\times$ more massive halos than those hosting obscured AGN, at $>5\sigma$ statistical significance, reiterating our result from \citetalias{cordova_rosado_cross-correlation_2024a}. In this analysis we find that unobscured AGN are also, on average, in $\sim 1.5\times$ more massive halos than their reddened counterparts (a $3\sigma$ difference). However, when considering solely the average central galaxy host halo mass, we reproduce our prior result that the unobscured AGN are found in halos that are $\sim 6\times$ more massive than the halos that host obscured AGN at $4\sigma$ significance. Now that we have recovered the same clustering amplitudes with a spectroscopic sample (\citetalias{cordova_rosado_cross-correlation_2024b}), we believe that the clustering difference cannot be ascribed to photometric redshift systematics. Having retired redshift, luminosity, and contamination systematic effects as possible sources for this clustering difference (see \S \ref{sec:robustness} for additional details), we suggest this difference in halo mass is representative of a key environmental difference between Type I and Type II AGN. Given their halo mass, reddened Type I AGN may be an intermediate population as past studies have suggested \citep{fawcett_striking_2023}, and are worth studying in greater detail in the future.

We caution against direct comparison of the absolute derived parameters from DM halo clustering models given their propensity to systematic shifts given subtle analysis choices \citep[see also ][]{aird_agn-galaxy-halo_2021}. In extending our analysis to a full HOD treatment, there are substantial changes to the method of estimating halo mass parameters that were not employed in our previous work. It is reassuring that moderately model-independent quantities such as the galaxy/quasar bias $b_g$ for the different AGN subsamples are consistent at the $1\sigma$-level between this work and our previous angular correlation function analysis (\citetalias{cordova_rosado_cross-correlation_2024a}). We have again found that the average halo mass is significantly higher for the unobscured AGN population than for the obscured AGN sample, now with an alternate modeling approach. Our inferred $\langle M_h \rangle_c$'s are consistent with our previous two-halo analysis.  We refer the reader to comparisons with prior linear clustering analyses in \citetalias{cordova_rosado_cross-correlation_2024a}, \citetalias{cordova_rosado_cross-correlation_2024b}, where we compared our results with the two-halo clustering results one obtains when splitting AGN based on X-ray, UV/optical, or IR selections. Often, studies have been constrained by the size of the sample, the availability of redshifts, or the control of systematics like sample completeness.

As described above, we have observed that the one-halo regimes of our measured $\omega(\theta)$'s for the different AGN subsamples are substantively different. Figure \ref{fig:power-law-fit} shows that the unobscured AGN have a significantly shallower slope at small scales ($s<1\, {\rm Mpc}$) than the correlation function of obscured AGN. This contrast has been previously observed in a variety of studies of Type I and Type II clustering (see discussion below). Given the shallowness of the unobscured AGN one-halo term, and our derived $f_{sat}$ from fitting it with an HOD, the most straightforward explanation is that Type I AGN are rarely found in satellites, and substantially less often than Type II AGN or LRGs with a similar bias ($f_{sat} \sim 10\%$). Our derived $f_{sat}$ posteriors for our unobscured, LRG, and obscured + reddened samples are shown in the right-hand panel of Figure \ref{fig:lit_fsat_comp}. The satellite fraction of obscured AGN ($\sim30\%$) is consistent for a sample of their characteristic average halo mass \citep[cf.][]{zehavi_galaxy_2011, krolewski_tomographic_2025}. For unobscured AGN, however, their inferred satellite fraction is significantly lower than the $f_{sat}$ of galaxies of comparable bias, as has been measured for LRG and un\textit{WISE} galaxy samples, see Table 2 of \cite{ishikawa_halo-model_2021}, and Table 3 of \cite{krolewski_tomographic_2025}. 

We are not the first to find a shallow inner slope for the correlation function of Type I AGN. Other analyses have also found that unobscured quasars have significantly weaker one-halo terms than typical galaxy samples. This is true for both optical \citep{villarroel_different_2014, jiang_differences_2016} and X-ray selections \citep{powell_swiftbat_2018, krumpe_spatial_2018}, suggesting that both classification methods are probing similar populations of unobscured quasars. Like us, these previous studies suggest that Type I AGN preferentially avoid satellite galaxies \citep{allevato_occupation_2012, richardson_halo_2012, shen_cross-correlation_2013, eftekharzadeh_halo_2019, krumpe_spatial_2023}. We compare the inferred satellite fractions from our and other HOD analyses of Type I AGN in the right panel of Figure \ref{fig:lit_fsat_comp}, alongside our inferred LRG and obscured + reddened AGN samples. We note that all estimates of $f_{sat}$ for Type I AGN are consistently below the expectation given by the posterior of LRGs or our obscured AGN catalogs. Prior Type II AGN analyses have only fitted restricted HOD models (with fixed parameters), which precludes an accurate $f_{sat}$ inference \citep[c.f.][]{krumpe_spatial_2018, powell_swiftbat_2018, petter_host_2023}. To our knowledge, this analysis of obscured AGN is the first of its kind to accurately estimate a Type II satellite fraction.

Our inferred AGN satellite fractions could have significant implications for the intra-halo environment in which Type I and II AGN occur. One possible avenue to explain the one-halo term differences, taken at face value, is that unobscured AGN are rarely triggered in satellite galaxies. This relative dearth is likely mediated by the assembly history of the systems, fuel availability at the center of the halo, possible boosts in black hole mass for a central galaxy, or an increased rate of merger events at the center of the gravitational potential. Taken together with the difference in average host halo mass $\langle M_h \rangle$, our results give us hints of a potential new description of the connection between the AGN halo environment and their spectral properties. While the satellite fraction that we measure for unobscured AGN is consistent with zero, we are only confident, based on the slope of the one-halo term, that they occur in substantially fewer satellite galaxies than their obscured analogs. 

We note, however, that unlike the foundation of standard HOD modeling tying galaxy luminosity functions to inferred halo masses \citep{van_den_bosch_cosmological_2013}, we have selected our AGN sample based on SMBH accretion properties that have not been shown to map neatly onto $M_h$ \citep{aird_agn-galaxy-halo_2021, powell_impact_2024}. Appropriate methods to model AGN clustering continue to be explored, whether it be through abundance matching techniques, semi-empirical modeling, or analytic methods like those used here \citep[cf. ][]{aird_agn-galaxy-halo_2021, powell_impact_2024}. Therefore, we are cautious in how we interpret these results beyond the comparisons that the empirical correlation functions immediately suggest. Though it is beyond the scope of this analysis, we suggest that analytic or semi-empirical models tying HOD parameters to the physics of SMBH growth should be explored to further understand how accretion physics impacts the systems' clustering. 

Our results show that the most MIR luminous Type I and Type II AGN occupy different-mass dark matter halos, on average. Additionally, Type I AGN are much less frequently found in satellite galaxies. Taken together, we believe these distinct observational results at a fixed MIR luminosity threshold cannot be explained by strict unified AGN morphological models \citep{urry_unified_1995, almeida_nuclear_2017}. These results suggest that AGN spectral classes may encode a spectrum of AGN triggering scenarios, which correlate with the halo environment of the host galaxy. While forward modeling tools and semi-empirical approaches will be essential to continue unraveling the formation and co-evolution history of SMBH and their host galaxies, we have shown that clustering statistics give us key insights into the relationship between accretion-derived emission properties and the dark matter around galaxies that host AGN. 

\subsection{Inferences from BH Scaling Relations}\label{sec:toy}

\begin{figure*}[t]
    \centering
    \includegraphics[width=1\linewidth]{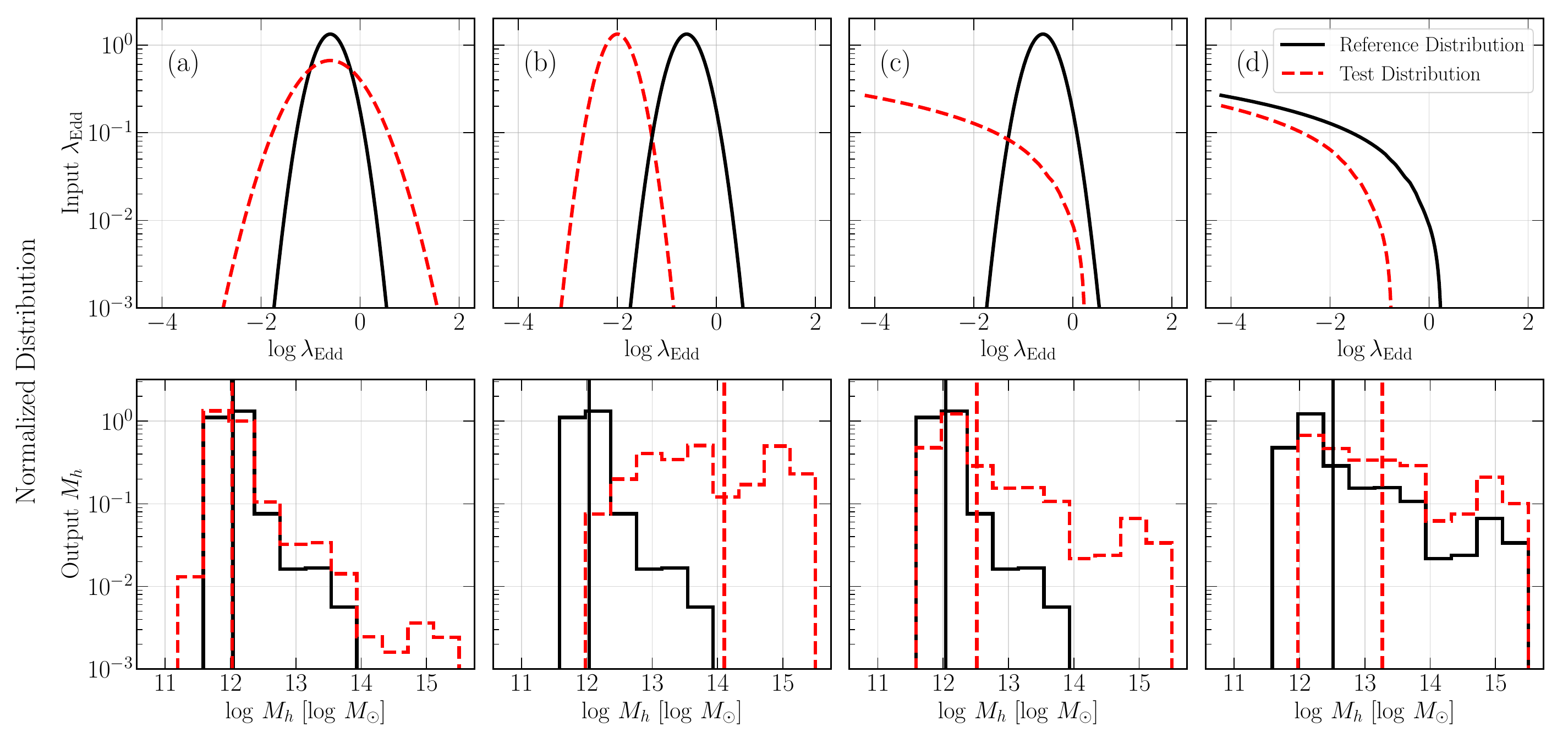}
    \caption{An illustration of our toy model investigating the effect of choosing different $L/L_{\rm Edd} \, (\lambda_{\rm Edd})$ distributions on the inferred halo masses from standard scaling relations. \textit{Top row}: Comparisons between previously posited $\lambda_{\rm Edd}$ distributions (black line) and variations of either the width, center, or functional form (dotted red line). We use the \citet{shankar_accretion-driven_2013} Gaussian $\lambda_{\rm Edd}$ distribution (centered at $\lambda_{\rm Edd} = 0.25$, $\sigma_{\log \lambda_{\rm Edd}} = 0.3$) as the reference in the first three columns. Column (a) provides a comparison with a broadened distribution ($\sigma_{\log \lambda_{\rm Edd}} = 0.6$), while the column (b) illustrates a re-centering to $\lambda_{\rm Edd} = 0.01$. Column (c) compares the \citet{shankar_accretion-driven_2013} $\lambda_{\rm Edd}$ with the broken power law proposed by \citet{jones_intrinsic_2016}, while column (d) compares the \citet{jones_intrinsic_2016} power law with a version of itself shifted by $\log \lambda_{\rm Edd} - 1$. \textit{Bottom row:} The output halo mass distributions for the upper $\lambda_{\rm Edd}$ distributions, following the scaling and $L_{bol}$ matching procedures described in \S \ref{sec:toy}. Vertical lines indicate the mean halo mass of the distribution. The $M_h$ distributions are sensitive to the shape of the input Eddington ratio distribution. We find that in order to substantially shift the average inferred halo mass from these scaling relations, the overall $\lambda_{\rm Edd}$ distribution must be shifted, as we do in the columns (b) and (d).} 
    \label{fig:toy_tests}
\end{figure*}

We consider what inherent physical differences between our AGN samples could be driving the difference in clustering we have found, having controlled for mid-IR luminosity, redshift, and sample contamination. We create a toy model using standard AGN scaling relations to test if Eddington ratio distributions could also affect inferred clustering properties. 

\subsubsection{Toy Model Definition}

Starting from galaxy stellar mass functions and our AGN clustering samples' matched luminosity distributions (detailed in \citetalias{cordova_rosado_cross-correlation_2024a}), we investigate what could feasibly reproduce the inferred halo difference we have measured. We build a simple model that takes standard scaling relations from the literature to turn bolometric luminosities to inferred halo masses, accounting for the empirical abundance of massive galaxies and SMBH's. The steps for this heuristic algorithm can be summarized as: 
\begin{enumerate}
    \item Define the black hole mass function at $z\sim1$ given a stellar mass function \citep[][in this case.]{van_der_burg_environmental_2013}. We convert the stellar masses to BH masses following the \citet{greene_intermediate-mass_2020} ``All'' galaxy sample relation. 
    \item Assume an $L_{bol}/L_{\rm Edd}$ distribution to convert BH mass to $L_{bol}$, assuming $M_{BH} \, [M_\odot] = L_{\rm Edd}/ \, 1.26 \times 10^{38} \, {\rm erg/s}$. 
    \item Sample from this simulated $L_{bol}$ distribution until it matches the empirical $L_{bol}$ distributions we infer for our high $L_{6\mu m}$ AGN samples. 
    \item Collect the associated $M_{BH}$ for the matched $L_{bol}$ distributions and translate them to $M_h$ following (and testing different) $M_{BH}-M_\star$  \citep[using][]{greene_intermediate-mass_2020} and $M_\star - M_h$ relations \citep[following][]{shuntov_cosmos2020_2022}. 
    \end{enumerate}

We will also compare the use of $M_{BH} - M_\star$ relations for early- (elliptical) versus late-type (spiral) galaxy relations \citep[cf. ][]{reines_relations_2015, davis_black_2018, greene_intermediate-mass_2020}, and relations derived from Type I quasars \citep{li_sloan_2023}. We assume a Gaussian $L/L_{\rm  Edd} = \lambda_{\rm Edd}$ distribution peaking at $\sim 0.25$ and $\sigma_{\log \lambda_{\rm Edd}} = 0.3$ to infer a $L_{\rm  Edd}$ distribution \citep[][the black line in the top panels of columns (a), (b), (c) of Figure \ref{fig:toy_tests}]{shankar_accretion-driven_2013}. Our empirical $L_{bol}$ distributions are produced by applying the \citet{hopkins_observational_2007} MIR bolometric correction to our measured $L_{6\mu m}$ distributions (shown in Figure 4 of \citetalias{cordova_rosado_cross-correlation_2024a}).

\subsubsection{Model Results}
Using the Type I and Type II AGN inferred clustering strengths and our empirical and consistent $L_{6\mu m}$ distributions as anchors in this analysis, we evaluate what change in our assumptions could produce the asymmetrical scaling necessary to make a difference in $M_h$. We consider if the MIR to bolometric correction could be different as a function of obscuration, finding that unobscured AGN would need to have a $\gtrsim 10\times$ larger MIR bolometric correction than obscured AGN to produce our inferred difference in $M_h$. However, recent studies have found the opposite -- that obscured sources may need $50\%$ higher bolometric corrections than unobscured sources \citep{duras_universal_2020}. Comparing different black hole to stellar mass scaling relations for distinct galaxy populations, their slight variations in $M_{BH} - M_\star$ are insufficient to produce the differences in halo mass we have inferred in our clustering analysis \citep[cf.][]{kormendy_coevolution_2013, reines_relations_2015, greene_megamaser_2016, saglia_sinfoni_2016, davis_black_2018, greene_intermediate-mass_2020}. Also, \citet{shankar_black_2019} \& \citet{li_sloan_2023} have shown that Type I AGN have more consistent stellar masses at fixed $M_{BH}$ when compared with quiescent galaxies. Additionally, there are few constraints in the literature at present that would suggest different kinds of AGN or galaxies require different prescriptions tying stellar mass to halo mass. There is also no clear evidence of a different stellar-to-halo mass relation for centrals and satellites \citep{watson_strikingly_2013, engler_distinct_2021}. 

We therefore test in more detail how changing the width, center, or parametrization of the $\lambda_{Edd}$ distribution affects our estimated halo mass distributions. The inferred $M_h$ distributions are shown in the bottom row of Figure \ref{fig:toy_tests}. We use the \citet{shankar_accretion-driven_2013} distribution as a reference, and note it leads to an average halo mass of $\sim10^{12} M_\odot$ in this model.  Widening our reference $\lambda_{\rm Edd}$ distribution to have $\sigma_{\log \lambda_{\rm Edd}} = 0.6$ serves to broaden the distribution of inferred $M_h$, without changing the average of $\sim10^{12.0} M_\odot$ (column (a) of Figure \ref{fig:toy_tests}). In column (b) Figure \ref{fig:toy_tests}, we show how shifting the center of \citet{shankar_accretion-driven_2013} distribution to $\lambda_{\rm Edd} = 0.01$ leads to inferring an average $M_h \sim10^{14.1} M_\odot$. Using empirical $\lambda_{\rm Edd}$ functional forms, such as the broken power-law distribution suggested by \citet{jones_intrinsic_2016} (column (c) of Figure \ref{fig:toy_tests}), leads to an average $M_h \sim 10^{12.5}$. When we shift the \citet{jones_intrinsic_2016} distribution by $\lambda_{\rm Edd}-1$ in column (d) of Figure \ref{fig:toy_tests}, we find a mean $M_h \sim 10^{13.2}$. These tests illustrate that, given consistent luminosity distributions and scaling relations, a possible avenue for distinct AGN samples to have different average $M_h$ is to have differentiated $\lambda_{\rm Edd}$ distributions.

Given this simple test, the much higher $M_h$ implies higher $M_{BH}$ for Type I's, and thus a lower Eddington ratio ($\lambda_{\rm Edd} \sim 3\%$) for the bulk of the Type I sample. Similarly, \cite{allevato_xmm-newton_2011} find Eddington ratio distributions of $\lambda_{\rm Edd} \gtrsim 0.01$ are required to match the observed halo mass and luminosity distributions for Type I AGN. Such a low typical $\lambda_{\rm Edd}$ is substantially at odds with single-epoch BH mass distribution estimates from broad line quasars, which are significantly broader and have $\langle  \lambda_{\rm Edd} \rangle \sim 0.05 - 0.25$ \citep[e.g.,][]{kollmeier_black_2006, kelly_constraints_2010} \citep[see also, ][]{marconi_local_2004, kelly_mass_2012, heckman_coevolution_2014}. However, this is a toy model. Further study connecting accretion properties to the dark matter environment in which AGN are found is crucial to build a more robust analysis.

\subsubsection{Future Directions}

Exploring the above scaling relations through semi-empirical modeling tools like \texttt{UniverseMachine} \citep{behroozi_universemachine_2019} and \texttt{TRINITY} \citep{zhang_trinity_2023}, as well as hydrodynamical simulations such as those used in recent studies like \citet{chowdhary_halo_2025}, are key avenues to pursue in future work. We can also use simulations to produce mock AGN catalogs that reflect accretion property distributions, and large upcoming spectroscopic data sets from DESI \citep{desi_collaboration_data_2025} and the Prime Focus Spectrograph \citep{greene_prime_2022}.

\section{Conclusions} \label{sec:conclu}

We present an angular cross-correlation study between luminous red galaxies observed by HSC and active galactic nuclei selected from HSC and \textit{WISE} photometry at scales of $0.02' < \theta < 200'$ ($0.01<s<100 {\rm \, Mpc}$ at $\langle z\rangle=0.9$). These AGN have been identified with a combination of HSC optical and \textit{WISE} MIR photometric colors, with a high spectroscopic verification rate presented in prior works \citep{hviding_spectroscopic_2024, cordova_rosado_cross-correlation_2024b}. Using three equatorial HSC fields totaling $\sim 600 \,{\rm deg^{2}}$, we calculated the correlation function from $1.5 \times 10^6$ LRGs and $\sim 28,500$ luminous AGN in the full redshift and luminosity range we probe ($z\in 0.7-1.0$, \, $L_{6\mu m} > 3\times 10^{44} {\rm \, erg\,s^{-1}}$). We perform a joint-likelihood fit of angular correlation functions across HSC fields with a 3-parameter halo occupation distribution, fitting physical scales $s > 0.1\, {\rm Mpc}$ ($\theta \gtrsim 0.2'$), and interpret the clustering amplitude and shape for the one- and two-halo components. Our principal conclusions are as follows.

\begin{enumerate}
    \item The host halos of unobscured (Type I) AGN ($\log \langle M_h \rangle = 13.43^{+0.02}_{-0.03} \, \log M_\odot$) are substantially ($\sim 3 \times$), and significantly ($>5\sigma$), more massive than the halos that host obscured (Type II) AGN ($\log \langle M_h \rangle = 12.91^{+0.07}_{-0.08} \, \log M_\odot$). Similarly, Type I quasars are hosted in halos that are $\sim1.5\times$ more massive than those that host reddened AGN (so-called ``red quasars'') ($\log \langle M_h \rangle = 13.27\pm0.05 \, \log M_\odot$), at $>3\sigma$. This difference is accentuated when selecting on the average central galaxy host halo mass, consistent with our prior findings. Having mitigated the possibility of systematic effects in photo-$z$ distributions or sample contamination with spectroscopic measurements, we are confident these results reflect real differences in the average host halo mass for luminous Type I and Type II AGN assuming this particular HOD formalism. 
    
    \item Having measured and modeled the intra-halo clustering from our AGN spectral sub-samples, we find that unobscured AGN have appreciably shallower one-halo terms than obscured AGN. We infer a satellite fraction from the HOD fit to these samples, finding that unobscured AGN have a characteristic $f_{sat} = 0.05^{+1}_{-0.05}\%$, with a mode of $1\%$, and a reddened + obscured AGN sample has $f_{sat} = 20^{+10}_{-5}\%$. This $>3\sigma$ difference suggests that Type I AGN are rarely found in satellite galaxies, indicating that some aspect of the intra-halo environment is correlated with obscuration level. Meanwhile, Type II AGN have a typical satellite fraction for galaxies of their bias, suggesting that their presence in satellites is proportional to their occupancy in galaxies overall.

    \item We measured substantive differences in the rates of halo occupancy for Type I and Type II AGN. Unobscured AGN are preferentially found in central galaxies with $M_h\sim10^{13.4} M_\odot$ ($M_{h,c}\sim10^{13.4} M_\odot$), while obscured AGN are triggered in galaxies with $M_{h}\sim10^{12.9} M_\odot$ ($M_{h,c}\sim10^{12.6} M_\odot$) and can be in either satellite or central galaxies. These results cannot be explained with strict unified (disk inclination-based) AGN models, and instead could suggest that AGN spectral classes are indicative of different phases or scenarios of SMBH accretion.

\end{enumerate}

In attempting to decipher the relationship between accreting SMBHs and the galaxies in which they reside, we have found that their halo environments -- as inferred from their clustering -- could vary significantly as a function of the AGN spectral class. Our results show that evolutionary models could be more capable of explaining the different types of galaxies in which AGN spectral types preferentially reside \citep[c.f. ][]{hopkins_cosmological_2008, alexander_what_2025}. Nevertheless, we stress that the HOD formalism is not tied to AGN accretion properties, and robust interpretations beyond the comparison between Type I and Type II will require more complex modeling choices tying HOD parameters to the physics of SMBH growth. Independent constraints of these properties from weak lensing measurements, X-ray selected samples, and analyses at a wider range of redshifts are clear extensions of this work and would help verify what we have measured here. Spectroscopic observing campaigns to build up the necessary number density to measure the one-halo term in a projected two-point function statistic will also supplement our current datasets. We have shown that large-scale clustering methods are a pivotal tool with which to disentangle SMBH and host galaxy properties. Analyses like ours will be substantially improved with data from upcoming wide-field and highly sensitive cosmological surveys \citep[eg. ][]{ivezic_lsst_2019, crill_spherex_2020, euclid_collaboration_euclid_2022, greene_prime_2022}. In so doing, we may arrive at a richer understanding of the SMBH-galaxy-halo connection, and the overarching role of black holes in our Universe.

\section*{Acknowledgments}

We thank C. Ward, P. Melchior, L. A. Perez, J. Givans, D. Setton, J. Myles, R. Wechsler, and A. Amon for helpful conversations throughout the course of this work. 

Computing was performed using the Princeton Research Computing resources at Princeton University. RCR acknowledges support from the Ford Foundation Predoctoral Fellowship from the National Academy of Sciences, Engineering, and Medicine. ADG and RCR gratefully acknowledge support from the NASA Astrophysics Data Analysis Program \#80NSSC23K0485. ADG and JEG acknowledge support from the National Science Foundation under Grant Number AST-1613744, and JEG acknowledges support from the National Science Foundation under Grant Number AST-2306950. AN acknowledges support from the European Research Council (ERC) under the European
Union’s Horizon 2020 research and innovation program with Grant agreement No. 101163128.

The Hyper Suprime-Cam (HSC) Collaboration includes the astronomical communities of Japan and Taiwan, and Princeton University. The HSC instrumentation and software were developed by the National Astronomical Observatory of Japan (NAOJ), the Kavli Institute for the Physics and Mathematics of the Universe (Kavli IPMU), the University of Tokyo, the High Energy Accelerator Research Organization (KEK), the Academia Sinica Institute for Astronomy and Astrophysics in Taiwan (ASIAA), and Princeton University. Funding was contributed by the FIRST program from the Japanese Cabinet Office, the Ministry of Education, Culture, Sports, Science and Technology (MEXT), the Japan Society for the Promotion of Science (JSPS), Japan Science and Technology Agency (JST), the Toray Science Foundation, NAOJ, Kavli IPMU, KEK, ASIAA, and Princeton University. 

This paper makes use of software developed for Vera C. Rubin Observatory. We thank the Rubin Observatory for making their code available as free software at \url{http://pipelines.lsst.io/}. This paper is based on data collected at the Subaru Telescope and retrieved from the HSC data archive system, which is operated by the Subaru Telescope and Astronomy Data Center (ADC) at NAOJ. Data analysis was in part carried out with the cooperation of Center for Computational Astrophysics (CfCA), NAOJ. 

We are honored and grateful for the opportunity of observing the Universe from Maunakea, which has the cultural, historical and natural significance in Hawai'i. 

This material is based upon work supported by the U.S. Department of Energy (DOE), Office of Science, Office of High-Energy Physics, under Contract No. DE–AC02–05CH11231, and by the National Energy Research Scientific Computing Center, a DOE Office of Science User Facility under the same contract. Additional support for DESI was provided by the U.S. National Science Foundation (NSF), Division of Astronomical Sciences under Contract No. AST-0950945 to the NSF’s National Optical-Infrared Astronomy Research Laboratory; the Science and Technologies Facilities Council of the United Kingdom; the Gordon and Betty Moore Foundation; the Heising-Simons Foundation; the French Alternative Energies and Atomic Energy Commission (CEA); the National Council of Science and Technology of Mexico (CONACYT); the Ministry of Science and Innovation of Spain (MICINN), and by the DESI Member Institutions: \url{https://www.desi.lbl.gov/collaborating-institutions}. Any opinions, findings, and conclusions or recommendations expressed in this material are those of the author(s) and do not necessarily reflect the views of the U.S. National Science Foundation, the U.S. Department of Energy, or any of the listed funding agencies. 

The authors are honored to be permitted to conduct scientific research on Iolkam Du’ag (Kitt Peak), a mountain with particular significance to the Tohono O’odham Nation.

\facilities{Subaru (HSC), WISE, NEOWISE, Mayall (DESI),
Sloan}

\software{\texttt{Astropy} \citep{astropy_collaboration_astropy_2018, astropy_collaboration_astropy_2022}, \texttt{Matplotlib} \citep{hunter_matplotlib_2007}, \texttt{NumPy} \citep{van_der_walt_numpy_2011, harris_array_2020}, \texttt{SciPy} \citep{virtanen_scipy_2020}}, \texttt{Corrfunc} \citep{sinha_corrfunc_2020}, Core Cosmology Library \citep{chisari_core_2019}

\bibliography{references-2, references, references2, references3}

\appendix

\section{Robustness of our Measurements and Inference} \label{sec:robustness}

\begin{figure}
    \centering
    \includegraphics[width=1.\linewidth]{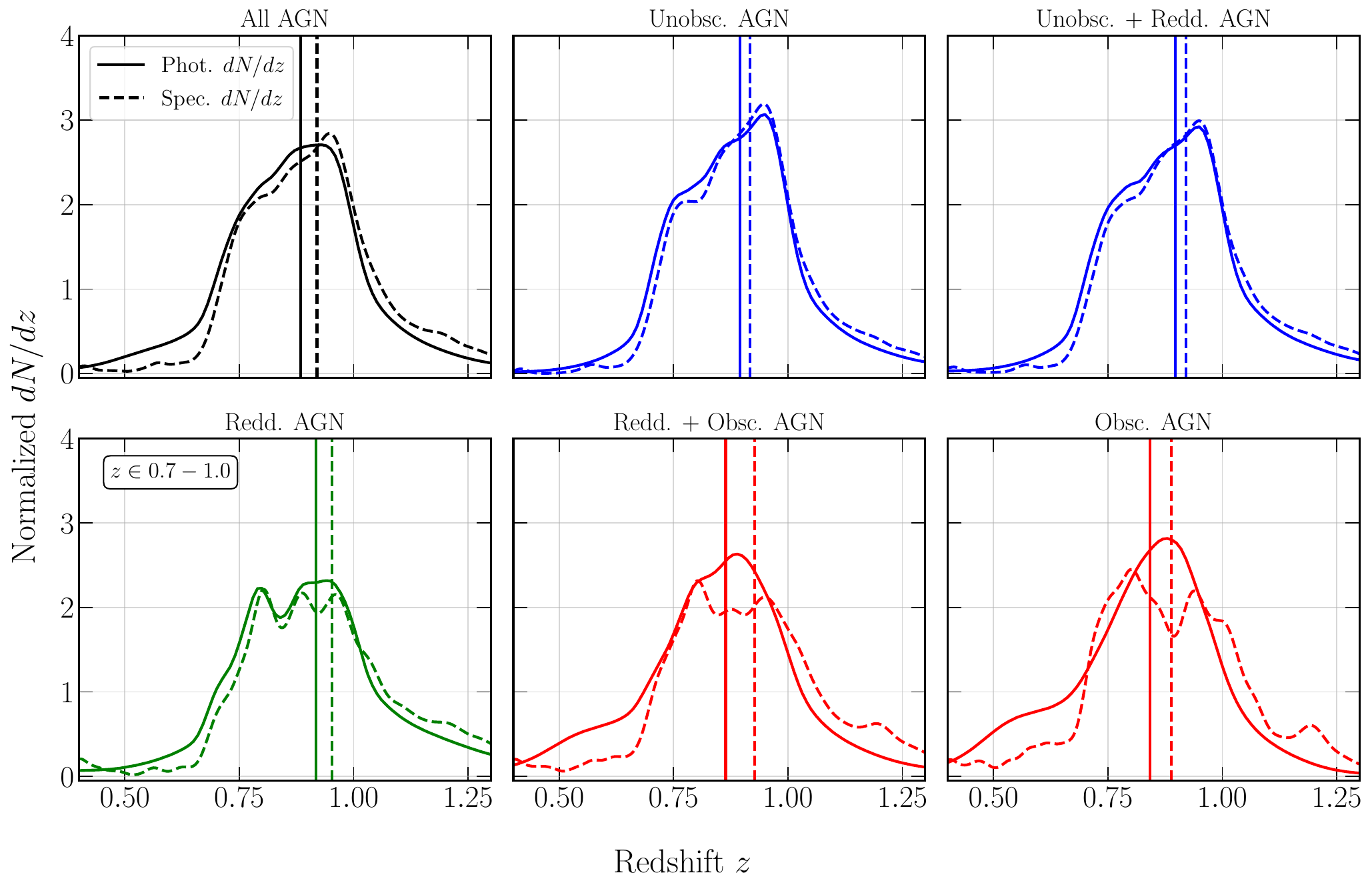}
    \caption{Normalized photometrically and spectroscopically-derived $dN/dz$ for each AGN sub-type sample, combing all HSC fields. Solid lines represent the weighted sum photometric $dN/dz$ for all objects across our three HSC fields with $L_{6\mu m} > 3\times 10^{44} \,{\rm erg/s}$ whose $p(z)$ is at least $3\%$ in our $z\in0.7-1.0$ redshift bin. We then cross-match these objects for any that have been targeted in DESI DR1, define a narrow Gaussian ($\sigma = 0.01\times (1+z)$) at each objects' measured spectroscopic redshift, and then sum them to make the dashed-line distributions. Vertical solid and dashed lines represent the median of each of these distributions; the shifts lie in the range $ 0.02 < \Delta z/(1+z) < 0.04$. Once propagated into our clustering interpretation, these shifts do not significantly change the inferred halo masses for our samples, nor the monotonic correlation between mass and obscuration we have inferred.}
    \label{fig:s_p_dndz}
\end{figure}

\begin{center}
    
\begin{table}
\centering
\caption{Measured Linear Bias and Inferred Halo Mass for Different Clustering Analysis Choices}
\begin{tabular}{c|cccccc}
\hline \hline
&  \multicolumn{6}{c}{AGN Samples} \\
{Parameter} & {All} & {Unobsc.} & {Unobsc.+Redd.} & {Redd.} & {Redd.+Obsc.}& {Obsc.} \\

\hline

& \multicolumn{6}{c}{Weighted Clustering and Photometric $dN/dz$ (Standard Approach)}\\

$b_g$ & $1.4\pm0.2$ & $2.3\pm0.2$ & $2.1\pm0.2$  & $1.6\pm0.2$ & $1.4\pm0.2$  & $1.2\pm0.2$ \\

$ \log M_{h}\,$[$\log h^{-1} \, M_\odot$]  &$ 13.2^{+0.1}_{-0.2}$ &  $13.4\pm0.1$ & $ 13.2^{+0.1}_{-0.2}$  & $ 13.0^{+0.1}_{-0.2}$  & $ 12.7\pm0.2$ & $ 12.6^{+0.2}_{-0.3}$  \\

\hline

& \multicolumn{6}{c}{Weighted Clustering and Spectroscopic $dN/dz$}\\

$b_g$ & $1.5\pm0.2$ & $2.4\pm0.2$ & $1.9\pm0.2$  & $1.8\pm0.2$ & $1.5\pm0.2$  & $1.2\pm0.2$ \\

$ \log M_{h}\,$[$\log h^{-1} \, M_\odot$]   &$ 13.2\pm0.1$ &  $13.4\pm0.1$ & $ 13.0^{+0.1}_{-0.2}$  & $ 13.0^{+0.1}_{-0.2}$  & $ 12.8\pm0.2$ & $ 12.6^{+0.2}_{-0.3}$  \\

\hline

& \multicolumn{6}{c}{Tomographic Clustering and Photometric $dN/dz$}\\

$b_g$ & $1.4\pm0.1$ & $2.3\pm0.1$ & $1.8\pm0.1$   & $1.4\pm0.2$  & $1.3\pm0.1$  & $1.3\pm0.1$ \\

$ \log M_{h}\,$[$\log h^{-1} \, M_\odot$]  &$ 12.9\pm0.1$ &  $13.4\pm0.1$ & $13.1\pm0.1$  & $12.7\pm0.2$   & $ 12.6\pm0.2$ & $ 12.7^{+0.1}_{-0.2}$  \\

\hline

& \multicolumn{6}{c}{Tomographic Clustering and Spectroscopic $dN/dz$ }\\

$b_g$ & $1.4\pm0.1$ & $2.3\pm0.1$ &  $1.8\pm0.1$  &$1.4\pm0.2$   & $1.3\pm0.1$  & $1.3\pm0.1$ \\

$ \log M_{h}\,$[$\log h^{-1} \, M_\odot$]  &$ 12.8\pm0.1$ &  $13.3\pm0.1$ &$13.0\pm0.1$ & $12.6\pm0.2$ & $ 12.5\pm0.2$ & $ 12.6^{+0.1}_{-0.2}$  \\

\hline \hline
\end{tabular}\label{tab:dndz_comp}
\vspace{3mm}

\end{table}

\end{center}

\subsection{Understanding the Type I AGN $M_1$ Posterior}
We consider if our measurement of an atypically low $f_{sat}$ for the unobscured AGN sample could be driven by fitting degeneracies of some kind. The MCMC-derived fit could be finding a solution that fits the model with high $M_1$ values without fully exploring its degeneracies with the $\alpha$ and $M_{min}$ parameters. To that end, we examine what $f_{sat}$ values are derived when we require the $M_1$ parameter avoid the edge of our prior. We test what HOD parameters are preferred by the unobscured AGN sample when the reddened AGN sample's $M_1$ $\pm 3\sigma$ posterior is used as a uniform prior on the same parameter for the unobscured sample. The inferred $f_{sat}$ value is again consistent with zero, and $>3\sigma$ inconsistent with the reddened AGN sample's $f_{sat}$ derived posterior. We additionally test how the AGN sub-samples HOD's behave when we fix $\log M_{min} = 12.5 \, \log M_\odot$ (the best-fit parameter for the full AGN sample). Here we again find that the inferred $\langle M_{h}\rangle$ for the Type I AGN is significantly higher than, and the $f_{sat}$ significantly lower than, the obscured AGN derived parameters. 

\subsection{Alternate HOD Models}
Next, we considered if the satellite fraction difference we found could be constrained irrespective of the adopted HOD formalism. We find the values of the inferred $f_{sat}$ change substantively when we apply our 5-parameter HOD -- Equations \eqref{eq:nc5}, \eqref{eq:ns5} -- to fit the data. The satellite fraction for obscured AGN is $13^{+13}_{-4} \%$, for the obscured $+$ reddened AGN sample it is $13^{+5}_{-3}\%$, $f_{sat}= 11\pm4 \%$ for the reddened AGN, while for the unobscured AGN the recovered value is consistent with zero with an upper $1\sigma$ uncertainty of $0.5\%$ (and a mode of $\approx 0.5\%$ from a skewed distribution). These values suggest that while a more flexible HOD model can find lower satellite fractions for all the samples by modeling more of the non-linear behavior as part of the central galaxy distribution, the order and statistical significance of the differences in $f_{sat}$ between unobscured and obscured AGN remain unchanged.

We also compare our fitting results when we use an emission line galaxy (ELG) HOD model \cite{comparat_01_2015, gonzalez-perez_host_2018, alam_multitracer_2020}. This 8-parameter model includes ``quenching'' terms (a high-halo mass cutoff of AGN activity, analogous to ELG clustering results), which could be necessary to better represent the potentially more complex halo environment of AGN populations \citep[c.f. ][]{aird_agn-galaxy-halo_2021}.  We observe the same difference in halo mass and in satellite fraction for unobscured vs. obscured AGN samples, suggesting that our halo occupation interpretations of the clustering signal are consistent even with more freedom in the HOD model.

\subsection{Spectroscopic $dN/dz$ estimate with DESI DR1}\label{sec:app_DR1_z}

While we were previously concerned with the potential effects of systematic biases in the $dN/dz$ of different photometric samples (see \S 4.4 of \citetalias{cordova_rosado_cross-correlation_2024a}), we found in \citetalias{cordova_rosado_cross-correlation_2024b} that the measured clustering amplitudes remained the same when using a spectroscopic sample to measure the clustering and infer the halo masses. We revisit this question by matching the HSC+\textit{WISE} AGN selection to the recently released DESI DR1 dataset \citep{desi_collaboration_data_2025}. We then use these spectroscopic $dN/dz$ distributions to re-estimate the linear clustering properties of different AGN sub-type samples.

As we performed in \citetalias{cordova_rosado_cross-correlation_2024b}, we can match the HSC+\textit{WISE} AGN catalog with the spectroscopic catalog published by \cite{desi_collaboration_data_2025}. The DESI DR1 footprint contains the complete area of the HSC PDR3 fields used in this analysis. We specifically match those objects whose \textit{WISE} \textit{W2} and \textit{W3} band S/N are each greater than 3, to be consistent with the selection used in this analysis. We refer the reader to our previous discussion of the DESI survey and selection effects in \textsection 2.4 of \citetalias{cordova_rosado_cross-correlation_2024b}, and DESI discussions referenced therein \citep{desi_collaboration_desi_2016, zhou_preliminary_2020, desi_collaboration_early_2024}. Repeating the same $0.5{''}$ on-the-sky match, we find there are $\approx 35,000$ matched sources across the GAMA, VVDS, and XMM fields, such that there they have a number density of 62 deg$^{-2}$, and of these approximately $50\%$ have luminosities above our $L_{6\mu m} > 3\times 10^{44} \,{\rm erg/s}$ threshold. 

We identify the objects that fall in our analysis redshift bin ($z\in0.7-1.0$) to construct the per-subtype $dN/dz$ shown in Figure \ref{fig:s_p_dndz} as solid lines. These are selected such that $\geq 3\%$ of each $p(z)$ is in the redshift bin. We then appropriately weight the $p(z)$ in the $dN/dz$ sum and in the clustering analysis (see Appendix B of \citetalias{cordova_rosado_cross-correlation_2024a} for a detailed description). For the clustering analysis, we build the $dN/dz$ for each field individually, as in our standard approach. We identify the subset of these objects that have a match in DESI DR1, and define a narrow Gaussian ($\sigma = 0.01\times (1+z)$) at each objects' measured spectroscopic redshift, and add these Gaussians up (with the appropriate weight from the photometric $p(z)$'s) to build the spectroscopic $dN/dz$ used in this comparison (dashed lines in Figure \ref{fig:s_p_dndz}). Comparing the median $z$ for the different $dN/dz$ across subtypes, there is a common shift across all sub-samples that lie in the range $ 0.02 < \Delta z/(1+z) < 0.04$. We replace the spectroscopic-match $dN/dz$ in our linear clustering analysis of the two-halo term with a complete covariance matrix, as performed in \citetalias{cordova_rosado_cross-correlation_2024a}, and re-measure the galaxy bias and inferred halo mass for these AGN samples. The results of this test, in comparison with our results in \citetalias{cordova_rosado_cross-correlation_2024a} following the standard approach we have used throughout, is shown in Table \ref{tab:dndz_comp}. We find that our results are entirely consistent when using the spectroscopic $dN/dz$, with no changes to the robust $>3\sigma$ differences in halo mass and galaxy bias between unobscured and obscured AGN we presented. 

We next tested our weighted analysis against the more typical tomographic redshift bin analysis. Objects with photometric redshifts are included in the nominal redshift bin (here $z\in0.7-1.0$) if the object's $p(z)$ median is found within the redshift bin, and the full $p(z)$ is then included to sum up to the full $dN/dz$. We perform our linear analysis once again while using the tomographic method with our high-$L_{6\mu m}$ photometric catalog, and find no shifts in our baseline results by more than $2\sigma$. The significance of the clustering differences between unobscured and obscured AGN samples is $>5\sigma$. These results are shown in Table \ref{tab:dndz_comp}. We repeat the match with DESI DR1-observed objects to build a tomographic $dN/dz$ from spectroscopic redshifts, and again see that the inferred halo masses are entirely consistent with the inferences using a photometric $dN/dz$; the $>5\sigma$ significance remains. From these tests, we retire our concerns of unconstrained systematic redshift uncertainties being a potential driver of these measured clustering differences. We continue using the most conservative and representative sampling of the redshifts used for our sample: a weighted clustering and $dN/dz$ measurement for our full photometric catalog.

\subsection{Potential Sample Contamination Test}

We also investigate why the clustering signal from obscured AGN is like that of typical young star-forming galaxies, with $\log \, \langle M_h \rangle \sim 12.4 \log M_\odot$ and $f_{sat}\sim 30\%$ \citep[cf. ][]{zehavi_galaxy_2011}. We inquire whether this measurement could be driven by sample contamination. While we found a confirmation rate of $92\%$ for our DESI-matched HSC+\textit{WISE} AGN (including objects below our $L_{6\mu m}$ threshold), these DESI spectra only identified sources down to $i<23$, a magnitude brighter than our sample selection limits for this work. We test what is the necessary contamination level and contaminating sample galaxy bias to depress our ansatz -- the unobscured AGN two-halo term -- such that it matches the amplitude of the obscured AGN two-halo term. We verify that if Type II AGN inherently share the same average halo mass of Type I AGN, then our obscured AGN would need $\sim50\%$ of the sample to be galaxies with halos of $M_h\sim 10^{8.2} M_\odot$, or $65\%$ would need to be galaxies with typical halos of $M_h\sim 10^{11.6} M_\odot$. We find these scenarios to be highly unlikely given prior spectroscopic confirmations, and infer that it is unlikely that the distribution of halos that host our obscured AGN are drawn from the same distribution as our unobscured AGN.

\subsection{$6\mu m $ Luminosity Threshold Tests}

Moreover, we also analyzed the shape of the cross-correlations when analyzing the clustering of AGN below our $L_{6\mu m}  = 3\times 10^{44} \, {\rm erg \, s^{-1}}$ threshold. We also consider if $L_{6\mu m}$ is an effective proxy for bolometric luminosity of an accreting SMBH, and whether bolometric corrections are relatively insensitive to the spectral type of the AGN. When comparing lower-luminosity AGN correlations with our fiducial results, we find that the amplitude of the two-halo terms are consistent for a given AGN spectral type, and the shapes of the one-halo term are also consistent (unobscured AGN flatten while obscured AGN continue to rise at small angular scales). This result suggests that even if there were a need for subtle bolometric corrections as a function of AGN spectral type, as \cite{duras_universal_2020} argue, the impacts on the clustering measurement are minimal.

Finding we have controlled for the samples' luminosity, completeness, contamination, and redshift uncertainty, we conclude that these correlation functions are robust and representative of luminosity-limited optical/MIR-selected and classified AGN. 

\end{document}